\documentclass[]{aa}

\usepackage{times}
\usepackage{graphics}
\usepackage{epsf}
\begin{document}

%\thesaurus{08    % Diffuse matter in space
%(
%09.03.1;      % ISM: clouds
%09.13.2;      % ISM: molecules
%13.09.3;      % Infrared: ISM : continuum
%13.19.3;      % Radio lines: ISM
%02.18.7;      % Radiative transfer
%09.09.1 L183  % ISM: Individual Objects
%) }

\title{Far-infrared and molecular line observations of Lynds 183 - 
studies of cold gas and dust
\thanks{Based on observations with ISO, an ESA project with instruments funded
by ESA Member States (especially the PI countries: France, Germany, the
Netherlands and the United Kingdom) and with the participation of ISAS and NASA.
}
}

\titlerunning{Far-infrared and molecular line observations of Lynds L183}

\author{
M.~Juvela \inst{1}
\and K.~Mattila \inst{2}
\and K.~Lehtinen \inst{2}
\and D.~Lemke \inst{3}
\and R.~Laureijs \inst{4}
\and T.~Prusti \inst{5}
}

\offprints{M.~Juvela}

\institute{
$^1$Helsinki University Observatory, T\"ahtitorninm\"aki, P.O.Box 14, 
SF-00014 University of Helsinki, Finland (mjuvela@astro.helsinki.fi) \\
$^2$Helsinki University Observatory, T\"ahtitorninm\"aki, P.O.Box 14,
SF-00014 University of Helsinki, Finland \\
$^3$Max-Planck-Institut f\"ur Astronomie, K\"onigstuhl 17, D-69117 
Heidelberg, Germany \\
$^4$
ISO Data Centre, Astrophysics Division, Space Science Department of ESA,
Villafranca del Castillo, P.O. Box 50727, 28080 Madrid, Spain \\
$^5$
Astrophysics Division, Space Science Department of ESA, ESTEC, PO Box 299,
2200 AG Noordwijk, The Netherlands
}

\date{Received <date> ; accepted <date>}

\abstract{
We have mapped the dark cloud L183 in the far-infrared at
100$\mu$m and 200$\mu$m with the ISOPHOT photometer aboard the ISO satellite.
The observations make it possible for the first time to study the distribution
and properties of the large dust grains in L183 without confusion from smaller grains.
The observations show clear colour temperature variations which
are
likely to be caused by changes in the emission properties of the dust
particles. In the cloud core the far-infrared colour temperature drops below
12\,K. The data allow a new determination of the cloud mass and the mass
distribution based on dust emission. The estimated mass within a radius of
10$\arcmin$ from the cloud centre is 25\,$M_{\sun}$.
We have mapped the cloud in several molecular lines including DCO+(2--1) and
H$^{13}$CO+(1--0). These species are believed to be tracers of cold and dense
molecular material and we detect a strong anticorrelation between the DCO+
emission and the dust colour temperatures. In particular, the DCO+(2--1)
emission is not detected towards the maximum of the 100$\mu$m emission where
the colour temperature rises above 15\,K. The H$^{13}$CO+ emission follows
closely the DCO+ distribution but CO isotopes show strong emission even
towards the 100$\mu$m peak. Detailed comparison of the DCO+ and C$^{18}$O maps
shows sharp variations in the relative intensities of the species.
Morphologically the 200$\mu$m dust emission traces the distribution of dense
molecular material as seen e.g. in C$^{18}$O lines. A comparison with dust
column density shows, however, that C$^{18}$O is depleted by a factor of
$\sim$1.5 in the cloud core.
We present results of $R$- and $B$-band starcounts. The extinction is much
better correlated with the 200\,$\mu$m than with the 100\,$\mu$m emission.
Based on the 200\,$\mu$m correlation at low extinction values we deduce a
value of $\sim$17$^{\rm m}$ for the visual extinction towards the cloud centre
where no background stars are observed anymore.
\keywords{ISM: clouds -- ISM: molecules -- Infrared: ISM: continuum -- Radio lines: ISM
-- Radiative transfer -- ISM: Individual objects: L183, L134N
}
}

\maketitle

\section{Introduction}

The distribution of material in dense interstellar clouds can be traced by
either molecular line emission or the continuum emission from dust particles.
The analysis of line spectra is complicated by radiative transfer effects. The
emission of radiation depends on local properties like density and kinetic
temperature. Based on these parameters, the observed line intensities can be
predicted with radiative transfer modelling. In practice, the density
structure of the sources is, however, unknown and e.g. small scale density and
velocity variations can seriously affect the observed intensities. Secondly,
the ratio between the abundance of the tracer and the total gas mass is only
known approximately and it can change even within one source. The chemical
reactions depend on density and temperature (Millar et al. \cite{millar97};
Lee et al. \cite{lee96a}) and therefore the chemistry of a dark cloud cores
will differ from the chemistry in the surrounding warmer material.

Molecular abundances depend on the strength of the ultraviolet radiation field
via processes such as photodissociation. The radiation is weakened by dust
extinction and by the self-shielding of the molecules and this introduces
radial abundance gradients. Selective photo-dissociation can lead to strong
abundance variations between different CO isotopes and can even affect the
excitation of the molecules (Warin et al.\cite{warin96}) The abundances change
even as a function of time (Leung et al. 1984; Herbst \& Leung 1989, 1990;
Bettens et al. 1995) and the observed abundances have been used as a `chemical
clock' to determine the evolutionary stage of sources (e.g. Stahler
\cite{stahler84}; Lee et al. \cite{lee96b}). Observations of e.g. TMC-1 show
clear anti-correlation between carbon bearing and oxygen bearing molecules
(Pratap et al. \cite{pratap97}). This can be compared with models (e.g. Bergin
et al. \cite{bergin95}, \cite{bergin96}) which show that CS and long carbon
chain molecules are produced preferentially at early times, while SO abundance
rises only much later.

The role of dust grains in chemical evolution is important but not yet well
understood. In dark cores some molecules can freeze onto the dust grains and
this causes the under-abundance of the molecules in the gas phase. CO
abundances can decrease by more than a factor of ten (Gibb \& Little
\cite{gibb98}). In the cold cores, the abundance of deuterated species, e.g.
DCO+ are enhanced. This fact has been used to identify condensations in the
earliest stages of star formation, before the birth of protostars again
increases the temperatures (e.g. Loren et al. \cite{loren90}). The abundance
ratio [DCO+]/[HCO+] depends also on the abundance of free electrons and can
therefore also be used to determine this quantity (e.g. Gu\'elin et al.
\cite{guelin77}, \cite{guelin82}; Watson \cite{watson78}; Wootten
\cite{wootten82}; Anderson et al. \cite{anderson99}). The electron abundance
is an important parameter since, through ambipolar diffusion, it affects the
timescales of physical cloud evolution and eventually the rate of the star
formation (Caselli et al. \cite{caselli98}).

The analysis of far-infrared dust continuum observations is in principle more
straightforward. The emission is optically thin and the intensity directly
related to the dust column density. At wavelengths 100-200$\mu$m the radiation
is emitted mainly by the classical large grains, with a possible small
contribution at 100\,$\mu$m by the `very small grains'. The large grains
heated by the normal interstellar radiation field settle to a temperature of
$\sim$17\,K and the maximum of the spectral energy distribution is close to
200$\mu$m. The temperature of the small grains fluctuates and most of the
radiation is at shorter wavelengths ($\lambda<100\mu$m). In the model of
D\'esert et al. (\cite{desert90}) some 10\% of the infrared emission at
100$\mu$m comes from very small dust grains.

The large grains are well mixed with the gas phase (Bohlin et al.
\cite{bohlin78}) and the FIR emission can be used to trace the gas
distribution. The dust emission corresponding to a given column density varies
depending on the composition and temperature of the dust. The far-infrared
colour temperature has been observed to drop towards the centre of many dark
cores (e.g. Clark et al. \cite{clark91}; Laureijs et al.\cite{laureijs91}) and
even cirrus type clouds (Bernard et al. \cite{bernard99}). It is clear that
the variations cannot be entirely due to a reduced radiation field, i.e. the
observed colour temperature variations indicate a change in the grain
properties. In the cloud cores, the dust particles may grow through
coagulation, and in this case the associated change in the grain emissivity
could explain part of the colour temperature variations (e.g. Meny et al.
\cite{meny00}; Cambr\'esy et al. \cite{cambresy01}). Changes in the dust
properties cause uncertainties to the column density estimates. These are,
however, smaller than those associated with the analysis of optically thick
molecular lines.

Most of the listed problems can be addressed by comparing observations of
molecular line emission with far-infrared dust emission. In this paper we
present such a study of the dark cloud L183.
%% ###!!!
%%

\subsection{L183}

L183 (L134N) is a well known dark cloud belonging to a high latitude cloud
complex (Galactic coordinates ($l$, $b$)=(6.0, 36.7)). The distance of the
cloud has been estimated to be 100\,pc (Mattila \cite{mattila79}, Franco
\cite{franco89}) although a distance of 160\,pc has also been used
(Snell \cite{snell81a}). We adopt $d$=100\,pc. Extinction towards the centre
of L183 is at least $A_{\rm V}\sim$6 (Laureijs et al. \cite{laureijs91}) and
may exceed $A_{\rm V}=$10. The cloud core is surrounded by an extended low
extinction ($A_{\rm V}<$1) envelope.

L183 was observed in CO(1--0) and $^{13}$CO(1--0) by Snell (\cite{snell81}),
in NH$_3$ by Ungerechts et al. (\cite{ungerechts80}) and
in several lines including C$^{18}$O(1--0) and H$^{13}$CO+(1--0) by Swade
(\cite{swade89a}; \cite{swade89b}).
%% mapped the cloud in several transitions:
%% $^{13}$CO(1--0), C$^{18}$O(1--0), CS(2--1), H$^{13}$CO+(1--0),
%% SO($N,J$=2,3--1,2), NH$_3$(1,1) and C$_3$H$_2$(1$_{10}$--1$_{01}$).
The $^{13}$CO line map shows a cometary shape with the head of the globule
pointing north in equatorial coordinates ($\sim$galactic east). This is also
the densest part of the cloud. The cloud has a sharp edge towards the north,
but in the southern part the strong $^{13}$CO and especially the $^{12}$CO
emission continues in the tail that points towards L1780. Together, the region
containing clouds L134, L169, L183, L1780 is sometimes called the L134 complex.
% (see Fig.~\ref{fig:ext}). 
NH$_3$ and C$_3$H$_2$ emission is concentrated into
the head region and ammonia emission is restricted to a very narrow ridge
running from south to north. The usual centre position given for the cloud,
(RA=15$^h$51$^m$30$^s$ Dec=-2$\degr$43$\arcmin$31$\arcsec$, 1950.0) is close
to the centre of the ammonia emission where density reaches a few times
10$^4$\,cm$^{-3}$ (Swade
\cite{swade89b}). In recent years the cloud has been observed in a number of
molecular lines (see e.g. Swade \& Schloerb \cite{swade92}; Turner et al.
\cite{turner99}, \cite{turner00} and Dickens et al. \cite{dickens00} and
references therein).

The infrared emission from L183 was studied by Laureijs et al.
(\cite{laureijs91}; \cite{laureijs95}) based on the IRAS data. The 60$\mu$m
emission is detected only in a narrow layer surrounding the cloud core. Based
on the upper limit of the I(60$\mu$m)/I(100$\mu$m) ratio, an upper limit of
15\,K was derived for the colour temperature assuming a $\nu^2$ emissivity
law. However, at 60$\mu$m, most of the emission comes from small dust
particles, and the colour temperature is not directly related to the physical
grain temperature. In the cloud core the lack of 60$\mu$m emission was
interpreted as evidence for mantle growth. Ward-Thompson et al. (\cite{wt94};
\cite{wt00}) have made sub-millimetre continuum observations of the cloud and
have detected two point sources close to the cloud centre. Lehtinen et al.
(\cite{lehtinen00}) have studied the properties of the sources using
far-infrared observations made with ISO.

In this paper we study the dark cloud L183 using 100\,$\mu$m and 200\,$\mu$m
maps and new molecular line observations. The far-infrared observations were
carried out with the ISOPHOT instrument aboard the ISO satellite. Both
100$\mu$m and 200$\mu$m are dominated by emission from classical large grains
and this makes it possible, for the first time, to study the variations in the
properties of the large grains without confusion from smaller size dust
populations. The far-infrared observations enable us to determine the mass
distribution of L183 independently from the molecular line data which may be
affected by chemical fractionation and other systematic effects. The dust
distribution will be compared with molecular line observations which include
new $^{12}$CO, $^{13}$CO and C$^{18}$O observations that cover the entire
cloud L183. The correlations with dust column densities will be used to
determine the degree of C$^{18}$O depletion in the core of L183. New
DCO+(2--1) and H$^{13}$CO+(1--0) mappings have been made with higher spatial
resolution than what were previously available. DCO+ is known to be a tracer
of cold gas and we will determine the correlations between DCO+ and dust
emission, especially the FIR colour temperatures. We will also study the
variations in the intensity ratio of DCO+ and H$^{13}$CO+. Finally, we present
results of $R$- and $B$-band star counts and a comparison between the
extinction and the far-infrared emission.

\section{Observations and data processing} \label{sect:observations}

\subsection{FIR observations} \label{sect:irobs}

Far-infrared observation of L183 were done with the ISOPHOT photometer
(Lemke et al. \cite{lemke}) aboard the ISO satellite (Kessler et al.
\cite{kessler}). Two 30$\arcmin \times 30\arcmin$ maps were made in the
PHT22 staring raster map mode with filters C\_100 and C\_200. The reference
wavelengths of the filters are 100\,$\mu$m and 200\,$\mu$m and the
respective pixel sizes $\sim$44$\arcsec$ and 89$\arcsec$. 

Additional narrow strips running from NW towards the cloud centre were
observed at 80$\mu$m, 100$\mu$m, 120$\mu$m, 150$\mu$m and 200$\mu$m. The
analysis of these observations will be presented by Lehtinen et al.
(in preparation).

\begin{figure}
\resizebox{\hsize}{!}{\includegraphics{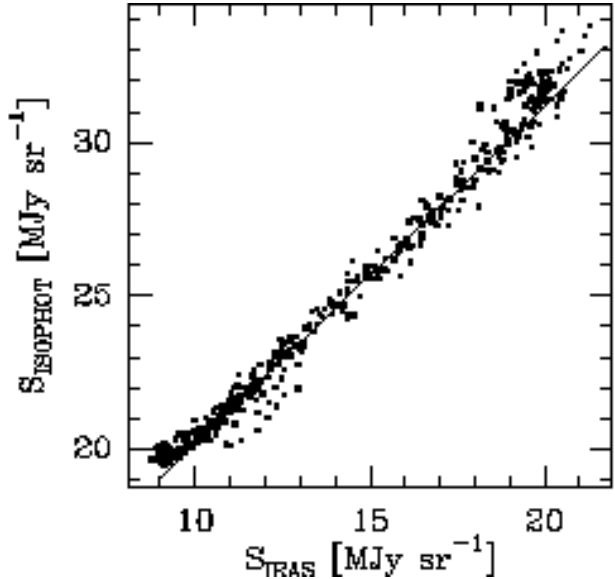}}
\caption[]{
The relation between the 100$\mu$m ISOPHOT and IRAS ISSA surface brightness
values. The ISOPHOT time-ordered data have been tilted 5\% in order to
obtain lowest dispersion when compared with IRAS values (see text). The slope
of the fitted line is 1.09. No zodiacal light has been subtracted from the
ISOPHOT values.}
\label{fig:iso_iras100}
\end{figure}

\begin{figure}
\resizebox{\hsize}{!}{\includegraphics{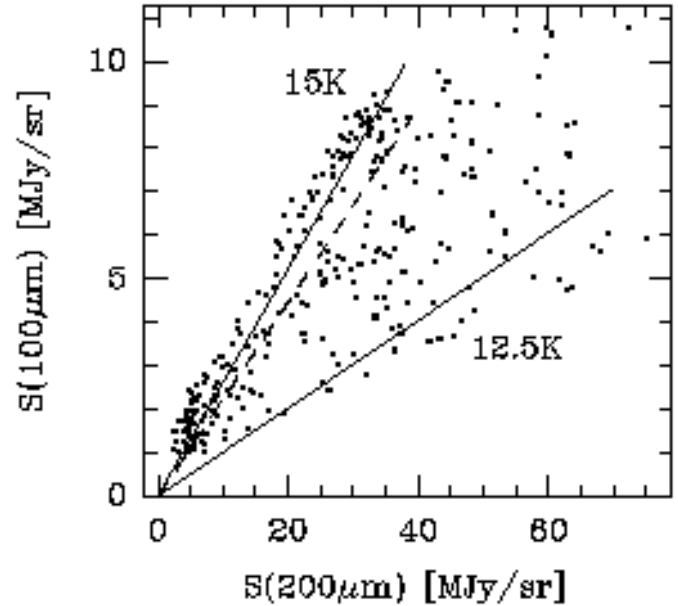}}
\caption[]{
Observed 100\,$\mu$m surface brightness as the function of the 200\,$\mu$m
values. The diffuse background determined from regions with lowest surface
brightness has been subtracted from the observed values and the 100\,$\mu$m
data are convolved to the resolution of the 200\,$\mu$m observations.. The
solid lines correspond to dust colour temperatures 12.5\,K and 15.0\,K. The
dashed line shows the least squares line fitted to points
$S$(200\,$\mu$m)$<$30\,MJy\,sr$^{-1}$ }
\label{fig:c100_c200}
\end{figure}

Comparison of the ISOPHOT 100\,$\mu$m observations with the 100\,$\mu$m IRAS
ISSA maps showed peculiar scatter with two separate branches in the ISOPHOT
vs. IRAS relation. The calibration of the ISOPHOT observations was done with
Fine Calibration Source (FCS) measurements that were done before and after the
actual mapping and a linear interpolation was performed to determine the detector
responsivity at each map position. The detector signals did not, however,
stabilize during FCS measurements and this leads to some uncertainty. A 5\%
change in the relative responsivities of the two FCS measurements was found to
be enough to minimize the scatter relative to the IRAS map. The average
responsivity over the map was not changed. The resulting relation between
ISOPHOT and IRAS ISSA surface brightness values is shown in
Fig.~\ref{fig:iso_iras100}. Because of the large temperature variations the
relation between 100$\mu$m and 200$\mu$m surfaces brightness values shows a
significantly larger scatter (see Fig.~\ref{fig:c100_c200}). This is also
decreased by the 5\% adjustment of the 100$\mu$m responsivities.
% and the scatter would reach minimum after a
% further 5\% tilting of the 100$\mu$m map. However, only the 5\% correction was
% applied to the 100\,$\mu$m data.

\begin{figure*}
\resizebox{\hsize}{!}{\includegraphics{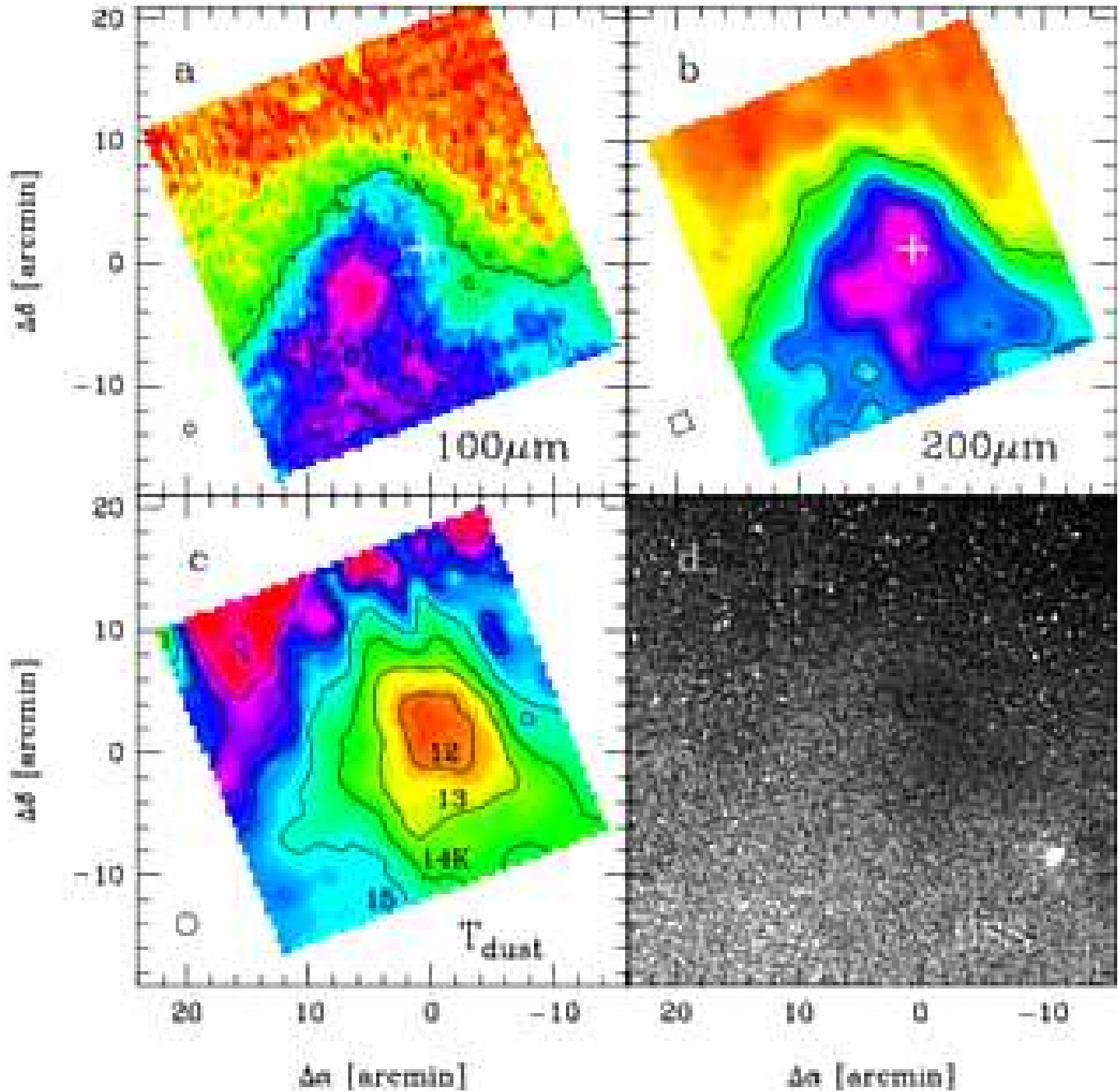}}
\caption[]{{\bf a-b)} The 100$\mu$m and 200$\mu$m ISOPHOT maps of the cloud L183.
The cross indicates the position of the continuum source detected by
Ward-Thompson et al. (\cite{wt94}. The (0,0) position is RA=15$^h$51$^m$30$^s$
DEC=-2$\degr$43$\arcmin$31$\arcsec$. In the case of 100\,$\mu$m the contours
are at 15 and 20\,MJy\,sr$^{-1}$ and for 200\,$\mu$m the contour levels are
30, 50 and 70 MJy\,sr$^{-1}$.
The small squares at the lower left corner indicate the pixel sizes,
44$\arcsec$ for 100\,$\mu$m and 89$\arcsec$ for the 200\,$\mu$m map.
{\bf c)} The map of dust colour temperature calculated from the 100$\mu$m and 200$\mu$m
ISOPHOT observations. The contours are at intervals of 1\,K starting with
12\,K. The effective resolution is 100$\arcsec$.
{\bf d)} blue image from the Digitized Sky Survey.
% ???
}
\label{fig:fir_maps}
\end{figure*}

The 100\,$\mu$m map was finally rescaled to the DIRBE surface brightness scale
by multiplying the values with 0.85. The scaling was established by
correlating DIRBE, IRAS and ISOPHOT surface brightness values in the area.
Details of the procedure are given in Lehtinen et al. (in preparation). The
200\,$\mu$m observations were calibrated using the FCS measurements which were
of good quality. The uncertainty in the absolute calibration of the
200\,$\mu$m map is expected to be below 20\% (see Garc\'ia-Lario
\cite{garcia00}). The final far-infrared maps are shown in
Fig.~\ref{fig:fir_maps}.

\subsection{Molecular line observations} \label{sect:molobs}

The molecular line observations were made with the SEST telescope during two
session in February 1998 and February 1999. During the first observing period,
the cloud was mapped in the lines $^{12}$CO(2--1), $^{13}$CO(1--0)
$^{13}$CO(2--1), C$^{34}$S(2--1) and C$^{18}$O(1--0). Due to a problem with
the receiver, the C$^{18}$O(2--1) map remained uncompleted. In February 1999
observations were made in the DCO+(2--1) and H$^{13}$CO+(1--0) lines.

During observations, the pointing was checked every two or three hours by
observing bright SiO maser sources. The estimated pointing uncertainty is less
than 5$\arcsec$. Most observations were made in the frequency switching mode
with a frequency throw 6\,MHz. The chopper-wheel method was used for
calibration and observed intensities are given in T$^*_{\rm A}$ units. The
typical rms noise levels are given in Table~\ref{table:observations}.

The $^{12}$CO, $^{13}$CO and C$^{18}$O maps cover practically the entire cloud
L183 as seen in optical extinction. In the southern part, a fairly strong
molecular line emission continues up to the edge of the map. The
H$^{13}$CO+(1--0) and DCO+(2--1) maps are smaller in size, but particularly in
the case of DCO+, they cover the whole region of significant emission. At the
map boundaries, the DCO+ intensity has dropped from a peak value $T_{\rm
A}\approx$1.1\,K to $\sim$0.25\,K or below.

\begin{table}
\caption[]{Summary of observed molecular emission lines. The columns are:
(1) transition, (2) beam size, (3) spectral resolution (4) number of observed positions
and (5) the average rms noise. }
\label{table:observations}
\begin{tabular}{rccrc}
line  &  FWHM       &  $\Delta v$     &  positions & $<\sigma_{\rm RMS}>$   \\
      & ($\arcsec$) &  (km\,s$^{-1}$) &            & (K)                  \\
\hline
$^{12}$CO(2--1)  & 23  & 0.054   &  83  &  0.17  \\
$^{13}$CO(1--0)  & 46  & 0.114   &  83  &  0.069 \\
$^{13}$CO(2--1)  & 24  & 0.057   & 532  &  0.14  \\
C$^{18}$O(1--0)  & 46  & 0.114   & 533  &  0.10  \\
%% C$^{18}$O(2--1)  & 24  & 0.057   &  50  &  0.22  \\
%% C$^{34}$S(2--1)  &    & 
H$^{13}$CO+(1--0) & 57  & 0.144   & 174  &  0.040 \\
DCO+(2--1)       & 35  & 0.087   & 176  &  0.061 \\
\end{tabular}
\end{table}

\subsection{Star counts}  \label{sect:counts}

The optical extinction in the L134 cloud complex was determined by means of
star counts using a blue (IIaO + GG385 filter) and a red (IIIaF + RG630) ESO
Schmidt plate (Plate Nos.3713 and 3714) which were obtained on 16 April 1980.
Star counts were performed using a reseau size of
2.8$\arcmin\times$2.8$\arcmin$. The total number of stars down to the limiting
magnitude of each plate was counted (see Bok \cite{bok56} for the method).

In order to derive the extinction, the function $N(m)$ is needed, i.e. the
number of stars per sq.degree with magnitude $\le m$ in a transparent
comparison area close to the dark nebula. The tables of van Rhijn
(\cite{rhijn29}) have often been used for this purpose (see Bok \cite{bok56}).
In the present case it was better to use the RGU photometric catalogue of
Becker and Fenkart (\cite{becker76}) for Selected Area 107 which is located
close ($l$=5.7 deg, $b$= 41.3 deg) to the L134 complex. The R and G band star
numbers for SA 107 were extracted from the compilation by Bahcall et al.
(\cite{bahcall85}). These $N(m)$ curves were extrapolated beyond the limiting
magnitudes $R$=17.5 and $G$=18.5, and were corrected to account for
the somewhat different Galactic latitude ($b$ = 36 deg) of the L134 complex by
using the tabulated results of Bahcall and Soneira (\cite{bahcall80}) for
their Galaxy model in the $B$ and $V$ bands.

An essential parameter of the $log N(m)$ vs. $m$ calibration curve is its
slope at the limiting magnitude of the star counts. In our case this slope
turned out to be 0.22 at $G_{\rm lim}$ = 20.1\,mag and $R_{\rm
lim}$=19.3\,mag, respectively. We note that the $G$ and $R$ passbands of the
Basle RGU photometric system with effective wavelengths of 463 and 638\,nm are
close enough to the wavelength ranges of our blue (380--480\,nm) and red plate
(630--690\,nm).

\begin{figure}
\resizebox{\hsize}{!}{\includegraphics{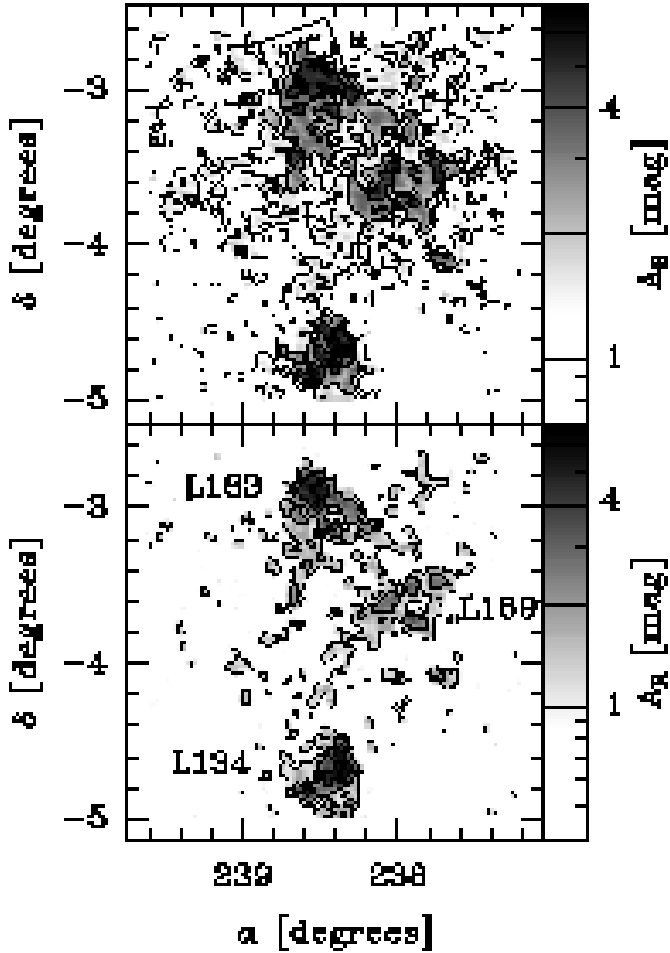}}
\caption[]{
The extinction maps of the L134 dark cloud complex derived from starcounts in
the $R$- (lower frame) and $B$-bands (upper frame). The area mapped with
ISOPHOT is indicated with a square. Individual clouds L134, L183 and L169 are
indicated in the figure. The contour levels are at 1, 2 and 4$^{\rm m}$.}
\label{fig:ext}
\end{figure}

The resulting extinction maps are shown in Fig.~\ref{fig:ext}.

\section{Observational results} \label{sec:results}

\subsection{FIR maps} \label{sec:iso_maps}

The main structure seen in the far-infrared maps (Fig.~\ref{fig:fir_maps}a-b)
is a wedge pointing northwards. The intensity is highest close to the head of
the wedge, and drops rapidly towards the north and more slowly towards the
south. There are clear differences in the distributions at 100$\mu$m and
200$\mu$m. There are two broad 200$\mu$m maxima, one close to the optical
extinction peak and previously known mm- and sub-mm continuum source
(Ward-Thompson et al. \cite{wt94}) and another $\sim$5$\arcmin$ towards SE. In
the 100$\mu$m map only the latter peak is visible. The relative strength of
the 200$\mu$m emission is larger in the north, west and southwest while the
100$\mu$m emission is stronger in the southeast. In the west the 100$\mu$m
emission seems to extend further out, although the surface brightness is
already close to the background level.

The 200$\mu$m emission is caused entirely by the large classical grains and at
100$\mu$m the contribution from small grains is expected to be small,
$\la$10\%. Dust temperature map was derived assuming that large grains emit
according to a modified black body radiation $\nu^2 B(\nu, T_{\rm d})$. The
background corresponding to the lowest surface brightness values within the
mapped area was first subtracted. Average 100$\mu$m surface brightness values
were calculated corresponding to each 200$\mu$m measurement. Values were colour
corrected and the dust temperature was obtained by the fitting of the modified
Planck curve. Since the colour correction depends weakly on the assumed
temperature the procedure was repeated once using the temperature derived from
the initial fit.

The dust temperature in Fig.~\ref{fig:fir_maps}c has a well defined minimum
close to the position of the continuum source. The minimum temperature in the
200$\mu$m core is below $\sim 12$\,K while in the north and west the
temperature rises rapidly to $\sim$16\,K. There is no temperature maximum at
the 100\,$\mu$m emission maximum and the position does not differ in any way
from its surroundings. The 5\% tilting of the 100$\mu$m map
(Sect.~\ref{sect:irobs}) changed the temperatures by up to $\sim$1\,K at the
eastern and western borders (the first and last scan lines in the map). As the
average responsivity was not changed the temperature in the centre, i.e. in
the cold core, was not affected.

The correlation between 100\,$\mu$m and 200\,$\mu$m surface brightness values
(Fig.~\ref{fig:c100_c200}) shows considerable scatter. The scatter was
somewhat reduced by the correction applied to the relative responsivity of the
100\,$\mu$m FCS measurements (see Sect.~\ref{sect:irobs}). A further
modification of the responsivities would have very little effect on the
scatter, which is caused by true variations of the dust properties. The
variations are reflected in the dust temperature distribution.

\begin{figure}
\resizebox{\hsize}{!}{\includegraphics{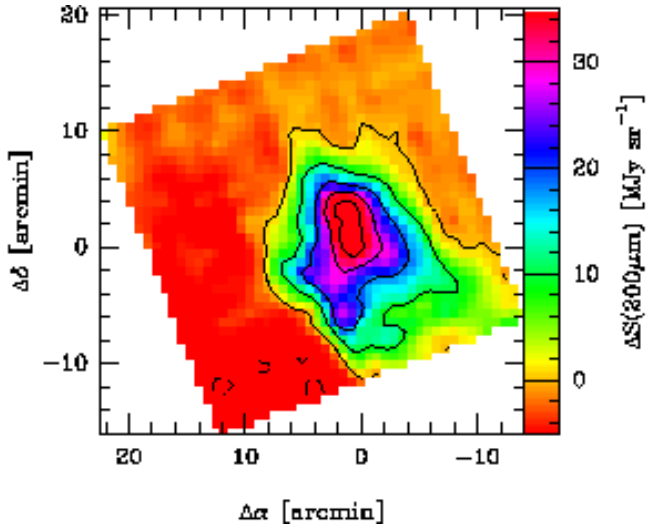}}
\caption[]{
Distribution of $\Delta S(200\,\mu$m). $\Delta S(200\,\mu$m) is the difference
between the observed 200\,$\mu$m surface brightness and the least squares
prediction calculated from the 100\,$\mu$m values. }
\label{fig:delta200}
\end{figure}

A linear least squares fit to the data points in Fig.~\ref{fig:c100_c200}
gives a relation $S_{\rm fit}$(200\,$\mu$m)= (0.14$\pm$0.01) S(100\,$\mu$m) +
(1.30$\pm$0.06). The fitting procedure takes into account the uncertainty
in both variables. The statistical error of the intercept in particular is
small compared with the uncertainty due to calibration and the background
subtraction. Data at positions where the background subtracted values of
$S$(200\,$\mu$m) were below 30\,MJy\,sr$^{-1}$ (i.e. outside the cloud core)
give a relation $S_{\rm fit}$(200\,$\mu$m)= (0.22$\pm$0.01) S(100\,$\mu$m).
The limit of 30\,MJy\,sr$^{-1}$ at 200\,$\mu$m corresponds approximately to
the 50\,MJy\,sr$^{-1}$ contour in Fig.~\ref{fig:fir_maps} where no background
subtraction was made. According to Fig.~\ref{fig:ext} the excluded area
corresponds also to the region with highest extinction, $A_{\rm B}\ga$4$^{\rm
m}$. 
Based on the derived dependence we calculate $\Delta
S(200\mu$m)=$S_{\rm obs}$(200\,$\mu$m)-$S_{\rm fit}$(200\,$\mu$m), which is
the difference of the observed 200\,$\mu$m surface brightness and the
predicted value based on the 100\,$\mu$m data. The resulting map of $\Delta
S$(200\,$\mu$m) (Fig.~\ref{fig:delta200}) shows the 200\,$\mu$m excess that
can be due to either cold dust or dust with enhanced emissivity at the longer
wavelength.

\begin{figure}
\resizebox{\hsize}{!}{\includegraphics{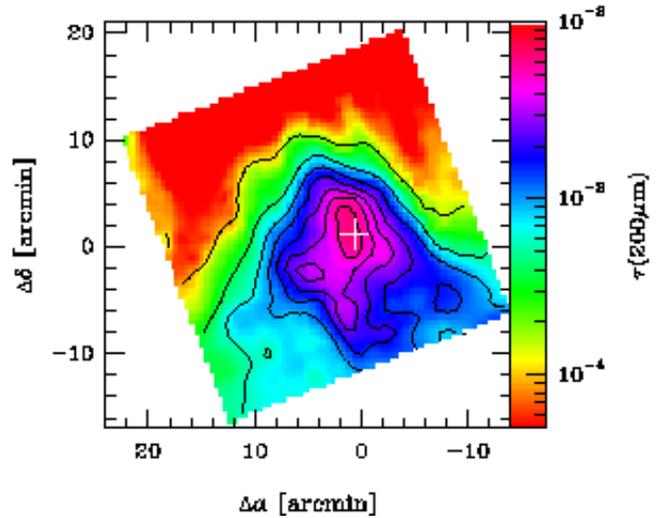}}
\caption[]{
200\,$\mu$m surface brightness (contours) and the calculated optical depths at
the same wavelength. The contours start at 15\,MJy\,sr$^{-1}$ and are drawn at
intervals of 15\,MJy\,sr$^{-1}$. }
\label{fig:c200_tau200}
\end{figure}

The 200\,$\mu$m optical depth was calculated from the observed intensities as
$\tau(200\,\mu$m)=$S/B(T)$. Here $B(T)$ is the black body intensity at
temperature $T$ as read from Fig.~\ref{fig:fir_maps}c. The resulting map,
which shows the column density distribution of large dust grains, is rather
similar to the 200\,$\mu$m intensity distribution. Due to its low colour
temperature, however, the main core is much more pronounced. The region of the
100\,$\mu$m peak is correspondingly weaker and is visible only as an extension
of the main core.
Fig.~\ref{fig:c200_tau200} shows the 200\,$\mu$m surface brightness in relation
to the 200\,$\mu$m optical depth. While the SE core is also visible as a separate
intensity peak in the 200\,$\mu$m map, the large grains are clearly
concentrated in the NW core.
A similar relation is seen between the 200\,$\mu$m emission and the dust
colour temperature. In the temperature map the SE peak is completely
invisible.

The cloud mass was calculated from the 200\,$\mu$m optical depth using the
value $\sigma_{\rm H}=2.5\cdot10^{-25}$cm$^{2}$ for the effective dust cross
section per hydrogen atom at 200\,$\mu$m (Lehtinen et al.~\cite{lehtinen98}).
The mass within 10$\arcmin$ of position (0,0) is 25\,$M_{\sun}$ for
$d=$100\,pc. Alternatively, the mass contained within the contour
$\tau$(200\,$\mu$m)=6$\cdot$10$^{-4}$ is 28\,$M_{\sun}$. The main uncertainty
is in the value of $\sigma_{\rm H}$ and use of the formula given by Boulanger et
al.~\cite{boulanger96}, $\sigma_{\rm H}=10^{-25} (\lambda/250\mu
m)^{-2}$cm$^{-2}$, would result in some 50\% higher mass estimates. We note
that the latter $\sigma_{\rm H}$ value is for diffuse medium, while the
Lehtinen et al. value is for a dense globule similar to L183.

As shown in Fig.~\ref{fig:delta200}, there is a clear 200$\mu$m excess in the
centre of L183. This could be due to a decrease in the physical temperature of
the dust grains induced by the decreasing intensity of the radiation field, or
by increased overall FIR emissivity of the grains. Another possibility would
be a change in grain properties, i.e. increased emissivity at the longer
wavelength, leading to a change of the dust emissivity index. Both explanations
are related to the increasing density in the cloud centre. Both theoretical
studies (e.g. Ossenkopf \cite{ossenkopf93}; Wright \cite{wright87}) and recent
PRONAOS observations (Bernard et al. \cite{bernard99}; Stepnik et al.
\cite{stepnik01}) indicate that dust emission properties are likely
to change in dense and cold clouds. This could be the result of grain growth,
or possibly the formation of grain aggregates. In a forthcoming paper (Juvela
et al., in preparation) we will study the far-infrared dust emission using
radiative transfer models. Further discussion about the dust temperatures and
changes in the dust emissivity law will be deferred to that paper.

\subsection{Distribution of molecular line emission} \label{sec:sest_maps}

\begin{figure}
\resizebox{\hsize}{!}{\includegraphics{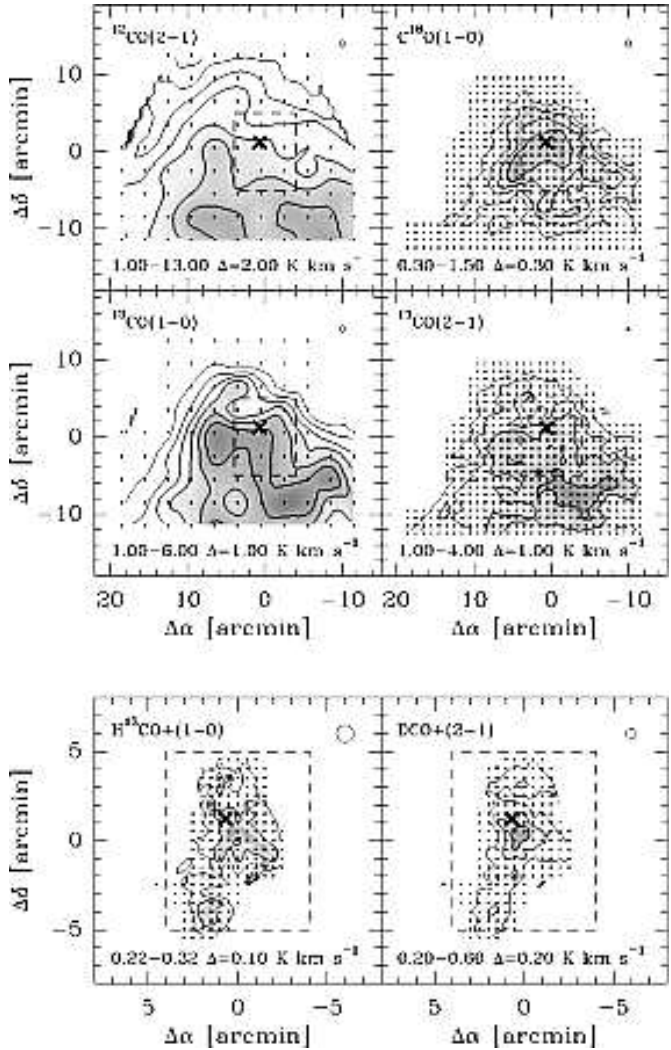}}
\caption[]{Maps of the integrated line areas:  $^{12}$CO(1--0),
C$^{18}$(1--0), $^{13}$CO(1--0), $^{13}$CO(2--1), H$^{13}$CO+(1--0) and
DCO+(2--1). The dashed lines in each frame indicate the location of the DCO+
core
}
\label{fig:map_all}
\end{figure}

Figure \ref{fig:map_all} shows the distributions of the integrated antenna
temperature in lines $^{12}$CO(2--1), $^{13}$CO(2--1), C$^{18}$O(1--0),
H$^{13}$CO+(1--0) and DCO+(2--1). 

The CO lines follow the general morphology seen in the maps by Snell
(\cite{snell81}) and Swade et al. (\cite{swade89a}). The $^{12}$CO emission
increases towards the south while C$^{18}$O and $^{13}$CO(1--0) delineate the
dense core close to the given centre position. The $^{13}$CO(1--0) emission
extends towards the southwest, however, where the $^{13}$CO(2--1) emission
peaks well outside the core seen in C$^{18}$O. The distributions of
H$^{13}$CO+(1--0) and DCO+(2--1) are clearly different from the CO species.
The distribution of H$^{13}$CO+(1--0) is similar to what was seen by Gu\'elin
et al. (\cite{guelin82}). However, due to the better resolution of the present
study the emission area is better resolved and the map is more structured.
There is a second emission peak south of the centre at position (1,-4). The
morphology of the DCO+(2--1) emission follows closely that of the
H$^{13}$CO+(1--0). The emission is concentrated along a narrow, but clearly
resolved ridge with some extension west of the centre position. The western
extension is seen in both lines, although it was not visible in the Gu\'elin
et al. H$^{13}$CO map. In the centre, the DCO+ emission seems to be more
concentrated around the peak position. However, this is mainly due to the
smaller beam size of the DCO+ observations.

\begin{figure}
\resizebox{\hsize}{!}{\includegraphics{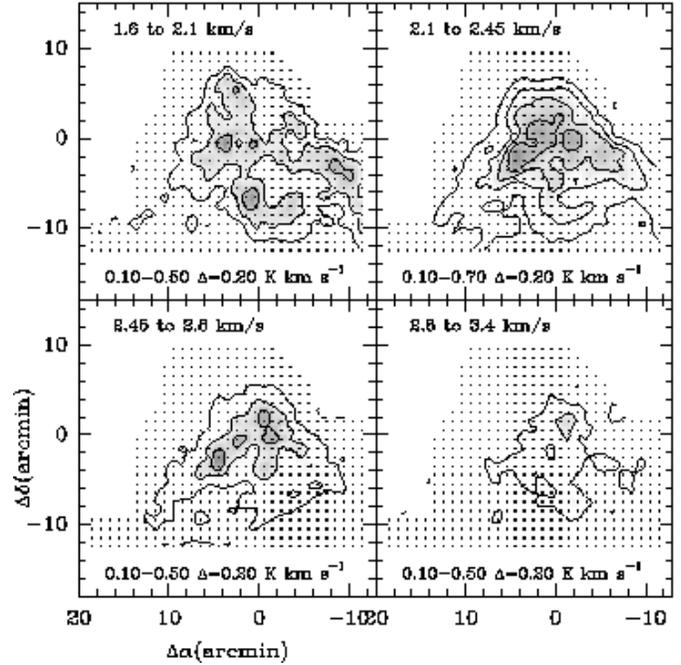}}
\caption[]{
C$^{18}$O(1--0) integrated antenna temperature in four velocity
intervals. The velocity intervals are given in the upper part of the panels.
The range of contour levels and the step between consecutive contours are
written at the bottom of the panels.
}
\label{fig:c18o10_channel}
\end{figure}

\begin{figure}
\resizebox{\hsize}{!}{\includegraphics{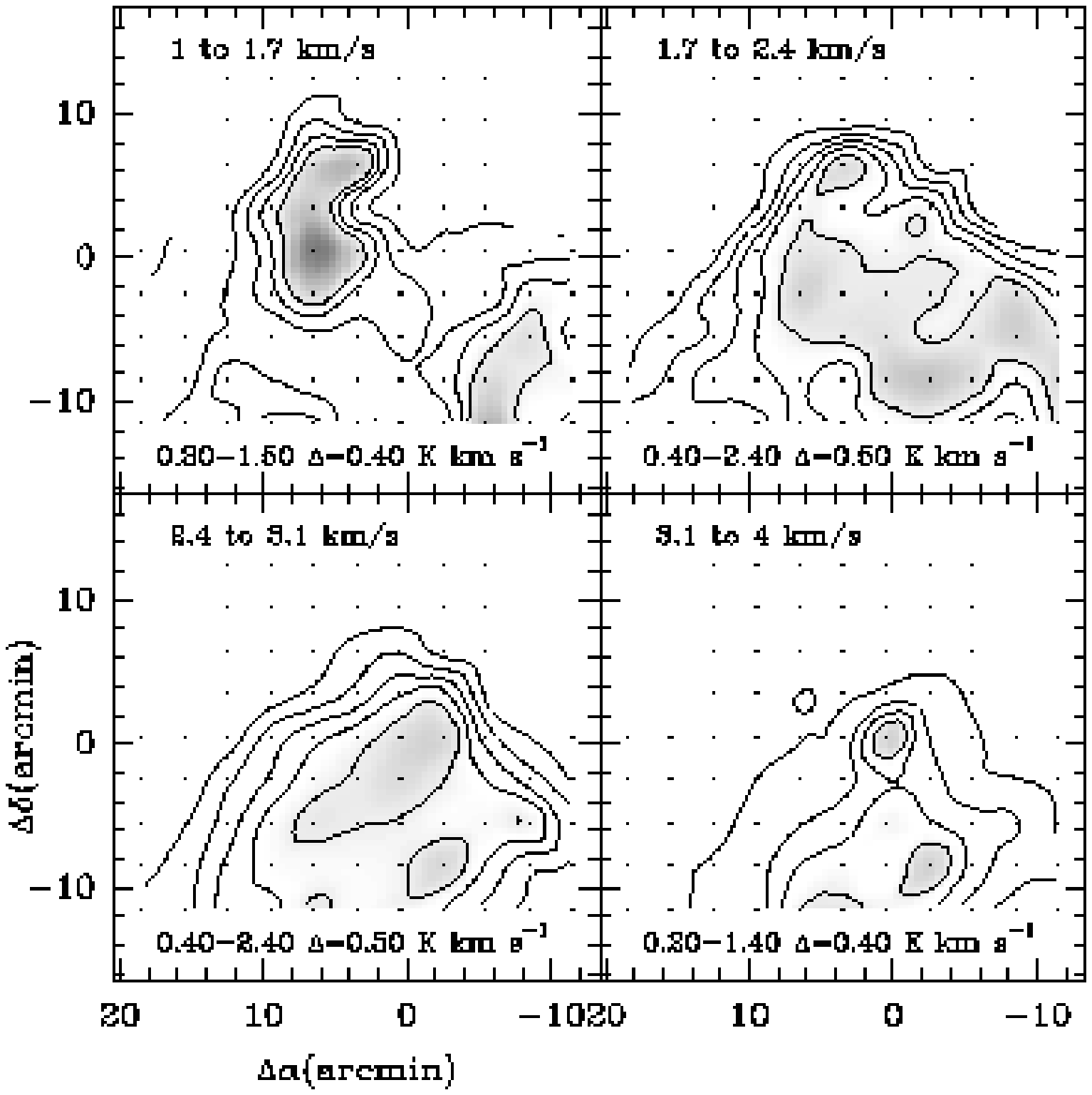}}
\caption[]{
$^{13}$CO(1--0) integrated antenna temperature in four velocity
intervals. The velocity intervals and the contour values are indicated in the
panels. }
\label{fig:13co10_channel}
\end{figure}

\begin{figure}
\resizebox{\hsize}{!}{\includegraphics{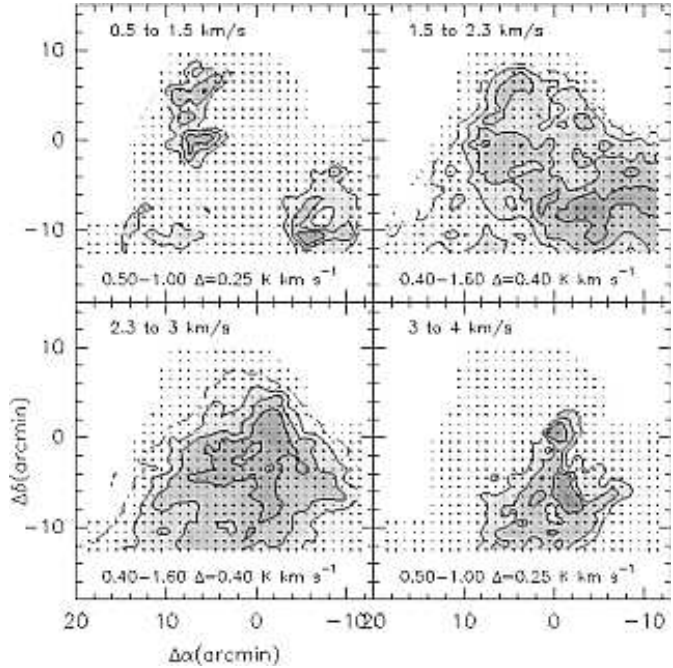}}
\caption[]{
$^{13}$CO(2--1) integrated antenna temperature in four velocity
intervals. The velocity intervals and the contour values are indicated in the
panels. }
\label{fig:13co21_channel}
\end{figure}

Maps of C$^{18}$O(1-0), $^{13}$CO(1-0) and $^{13}$CO(2-1) emission in
different velocity intervals are shown
Figs.~\ref{fig:c18o10_channel}--\ref{fig:13co21_channel}. In the lowest radial
velocity bin, C$^{18}$O emission is seen both northeast of the position (0,0)
and towards the western edge of the map. In the interval
~2.0-2.5\,km\,s$^{-1}$, strong emission still exists in the western part of
the map. The main core is visible in all three higher velocity intervals.
Below 3\,km\,s$^{-1}$ the emission extends between the positions of the
100\,$\mu$m and 200\,$\mu$m maxima (see Fig.~\ref{fig:fir_maps}) while at the
highest radial velocities the emission is concentrated around position (0,0)
only. The same general morphology and dependence on the radial velocity is
repeated in the $^{13}$CO maps. The relative intensity of the $^{13}$CO lines
is, however, higher in the south. As the radial velocity increases the
emission maximum moves over the C$^{18}$O core from northeast to southwest.

\begin{figure}
\resizebox{\hsize}{!}{\includegraphics{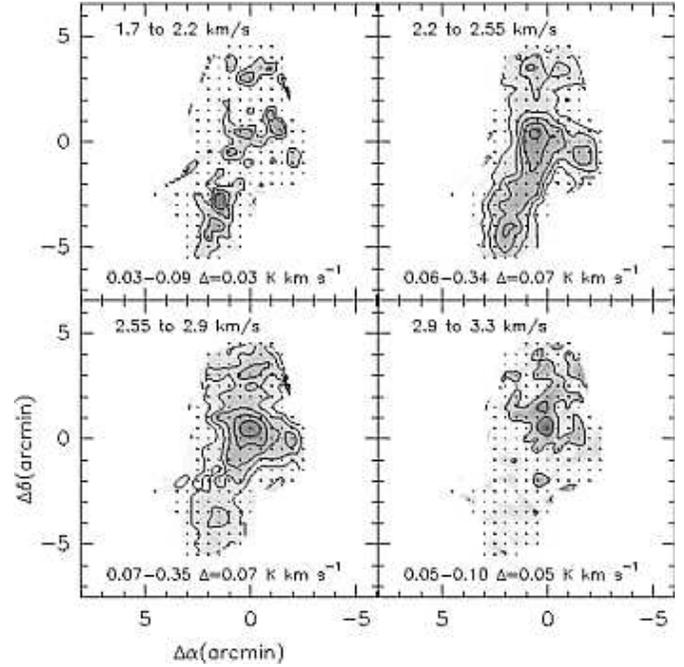}}
\caption[]{
DCO+(2--1) integrated antenna temperature. The velocity intervals and
the contour values are indicated in the panels. }
\label{fig:dco21_channel}
\end{figure}

\begin{figure}
\resizebox{\hsize}{!}{\includegraphics{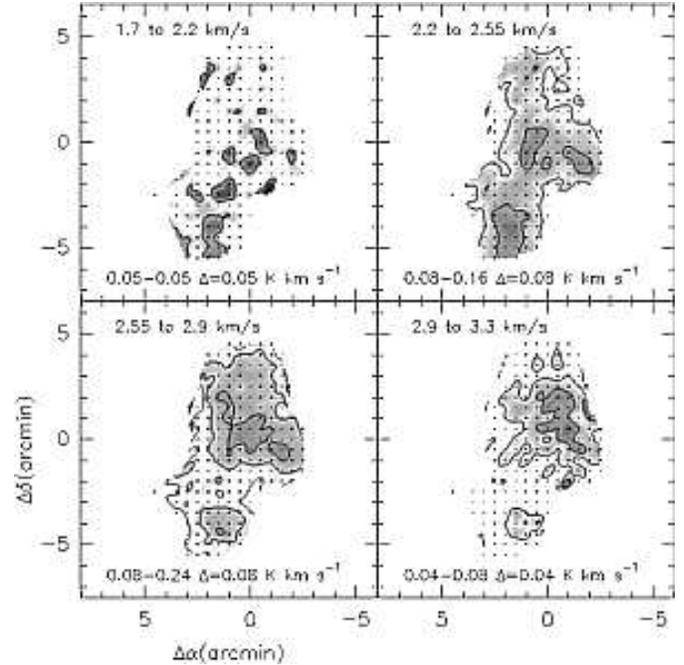}}
\caption[]{
H$^{13}$CO+(1--0) integrated antenna temperature. The velocity intervals and
the contour values are indicated in the panels. }
\label{fig:h13co10_channel}
\end{figure}

The DCO+(2--1) and H$^{13}$CO+(2--1) in
Figs.~\ref{fig:dco21_channel}-\ref{fig:h13co10_channel} show similar
behaviour. At low velocity, the emission is concentrated in the south around
position (2,-3) while at other velocities the maximum is reached close to
position (0,0). There are also some differences, and H$^{13}$CO+ is relatively
stronger around the emission peak in the south.

\section{Models of radiative transfer and column densities of molecular gas}
\label{sect:model}

\begin{figure}
\resizebox{\hsize}{!}{\includegraphics{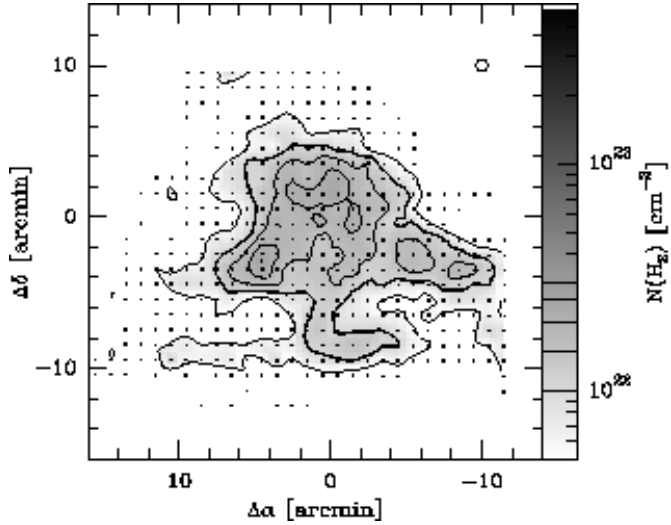}}
\caption[]{LTE hydrogen column density as derived from the
$^{13}$CO(1--0) and C$^{18}$O(1--0) observations. The contours start at value
N(H$_2$)=5$\cdot$10$^{21}$cm$^{-2}$ and are drawn at intervals of
5$\cdot$10$^{21}$cm$^{-2}$ (the thick contour line denotes the value 
N(H$_2$)=1$\cdot$10$^{22}$cm$^{-2}$). 
}
\label{fig:ph_colden}
\end{figure}

LTE column density estimates were computed based on the $^{13}$CO(1--0) and
C$^{18}$O(1--0) observations. The derived excitation temperatures are between
7 and 9\,K. Compared with the rest of the cloud, the excitation temperature
seems to be systematically ~1\,K lower in the northern part close to the
column density maximum. However, the difference is small and we derive column
densities assuming a constant excitation temperature of $T_{\rm ex}$=8.4\,K.
The LTE column densities were calculated according to the formula given by
Harjunp\"a\"a et al. (\cite{harjunpaa96}) and are shown in
Fig~\ref{fig:ph_colden}. In the conversion from C$^{18}$O to H$_2$ column
densities, a fractional C$^{18}$O abundance of 1$\cdot$10$^{-7}$ was assumed,
corresponding to the quiescent cloud positions in Harjunp\"a\"a et al.
(\cite{harjunpaa96}) and Frerking et al. (\cite{frerking82}).

From the derived column densities we obtain a value of $\sim$47\,M$_{\sun}$ for the
cloud mass within 10$\arcmin$ of the centre position, or some 35\,M$_{\sun}$
for the mass contained within the contour $N({\rm
H}_2$)=1.0$\cdot$10$^{22}$\,cm$^{-2}$.

For comparison, the C$^{18}$O and $^{13}$CO emission were modelled with
spherically-symmetric model clouds where the radiative transfer problem was
solved with Monte Carlo methods (see Juvela et al.~\cite{juvela97}). While
spherical symmetry presents an extreme simplification of true density
distribution, the models will still be more realistic than the assumptions
underlying the LTE calculations (see e.g. Padoan et al. \cite{padoan00}). The
model clouds are assumed to be isothermal with radial density dependence of
$n\sim r^{-1.5}$ and with density contrast 20 between cloud centre and outer
boundary. The free parameters of the models were the cloud size, central
density and the kinetic temperature. The C$^{18}$O(1--0), $^{13}$CO(1--0) and
$^{13}$CO(2--1) spectra within 10$\arcmin$ radius of the position (0,0) were
compared with spectra calculated from the radiative transfer model and the
free parameters were adjusted in order to obtain the best fit. The gas
distribution is asymmetric with respect to the centre position and the
$^{13}$CO emission is displaced with respect to C$^{18}$O. Therefore, the
results can not be expected to be very accurate. However, although the listed
free parameters are not well constrained, the column density estimates should
be more reliable.

Assuming a constant kinetic temperature of 10\,K, the best fit was obtained with
a model having a mass of 40\,$M_{\sun}$. The mass estimate is not very
sensitive to the assumed kinetic temperature. A similar model with $T_{\rm
kin}$ increasing linearly from 8.0\,K in the centre to 15\,K on the cloud
surface resulted in a mass estimate of 34\,$M_{\sun}$. The quality of the fit
was, however, marginally better in the isothermal model.

The spectral lines do not show signs of cloud collapse. Several $^{13}$CO
spectra have double peaked, asymmetric profiles but comparison with optically
thinner C$^{18}$O indicates that these are caused by separate emission
components rather than self-absorption. Gaussian fits give average linewidths
of $\sim$0.7\,km\,s$^{-1}$ for both $^{13}$CO(2--1) and C$^{18}$O(1--0). The
virial mass was estimated according to equation
\begin{equation}
M_{\rm vir} = (\frac{\sigma_{\rm 3D}}{C})^2 R,
\end{equation}
where $\sigma_{\rm 3D}$ is the three-dimensional velocity dispersion, $R$ the
radius of the cloud and $C$ is a constant depending on the assumed model of
the cloud. The value of $C$ is 0.0508\,km\,s$^{-1}$\,$\sqrt{pc/M_{\sun}}$ for
a homogeneous sphere and 0.0463\,km\,s$^{-1}$\,$\sqrt{pc/M_{\sun}}$ for
density distribution $n\sim r^{-1}$ (see Liljestr\"om \cite{liljestrom91}).
Assuming gas temperature 13\,K and $R\sim10\arcmin$ the observed velocity
dispersion corresponds to virial masses of 54\,$M_{\sun}$ and 44\,$M_{\sun}$
for the two models. In the presence of noise the FWHM derived from gaussian
fits may, however, be biased and the average FWHM read directly from the
spectra is smaller, $\sim$0.5\,km\,s$^{-1}$, lowering the virial mass
estimates to 36\,$M_{\sun}$ and 30\,$M_{\sun}$, correspondingly. The previous
analysis ignores both the external pressure and the presence of velocity
gradients. The cloud is, however, approximately in virial equilibrium. Even
though the cloud as a whole may not be collapsing it can still contain
collapsing fragments. In fact, there is evidence of several pre-protostellar
cores in L183 (Lehtinen et al.~\cite{lehtinen00}).

We will return to the modelling of the molecular line data in a future paper
(Juvela et al., in preparation) where radiative transfer models will also be
constructed for the far-infrared dust emission.

\subsection{Comparison of FIR and molecular line maps and column densities}

\begin{figure}
\resizebox{\hsize}{!}{\includegraphics{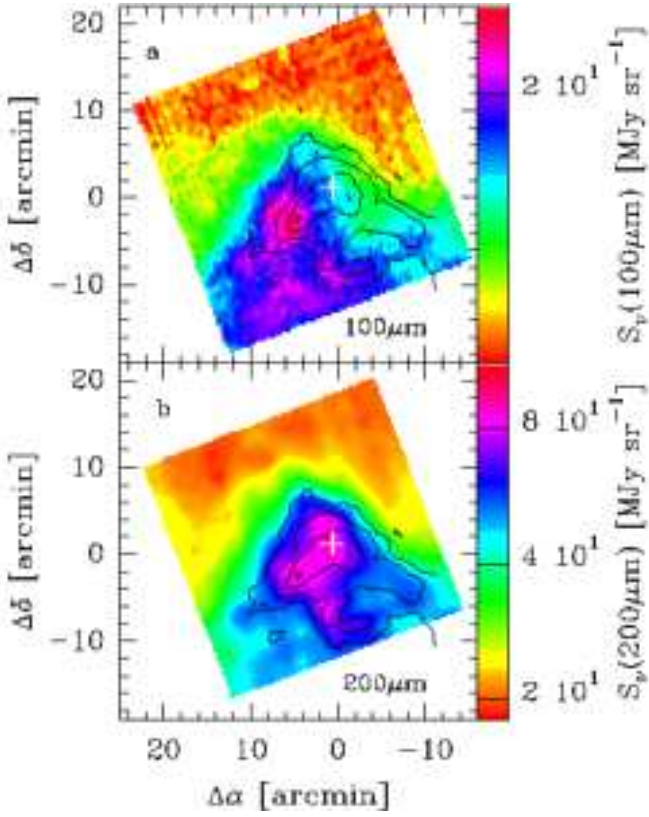}}
\caption[]{
The far-infrared maps of L183 at 100$\mu$m and 200$\mu$m with
C$^{18}$O(1--0) contours overlaid. The contours are drawn between
0.4\,K\,km\,s$^{-1}$ and 1.6\,K\,km\,s$^{-1}$ at intervals of
0.3\,K\,km\,s$^{-1}$}
\label{fig:iso_c18o}
\end{figure}

Fig.~\ref{fig:iso_c18o} shows the distribution of FIR surface brightness
relative to the intensity of the C$^{18}$O(1--0) lines. The 200$\mu$m
intensity closely follows the C$^{18}$O emission. The peak positions do
coincide and the second 200$\mu$m maximum in the southeast as well as fainter
extensions towards south and towards southwest have corresponding features in
the line map. The 200$\mu$m emission extends, however, somewhat further in the
north while the C$^{18}$O is relatively stronger in the southwest. With
respect to the 200\,$\mu$m peak the C$^{18}$O(1--0) maximum is displaced
slightly towards the southwest. Further towards the southwest, the C$^{18}$O
is again well traced by the FIR emission. The second C$^{18}$O(1--0) peak at
position (4,-2) coincides with the second 200\,$\mu$m peak. There is, however,
a third 200\,$\mu$m peak at (1,-6) which does not have a direct counterpart in
the molecular line emission. The peak corresponds to just one pixel above the
diffuse background and could even be caused by a single extragalactic source.
The 100$\mu$m peak traces the eastern extension of the C$^{18}$O distribution.
Apart from this there is a clear lack of correlation with molecular line
emission, especially within the C$^{18}$O core.

\begin{figure}
\resizebox{\hsize}{!}{\includegraphics{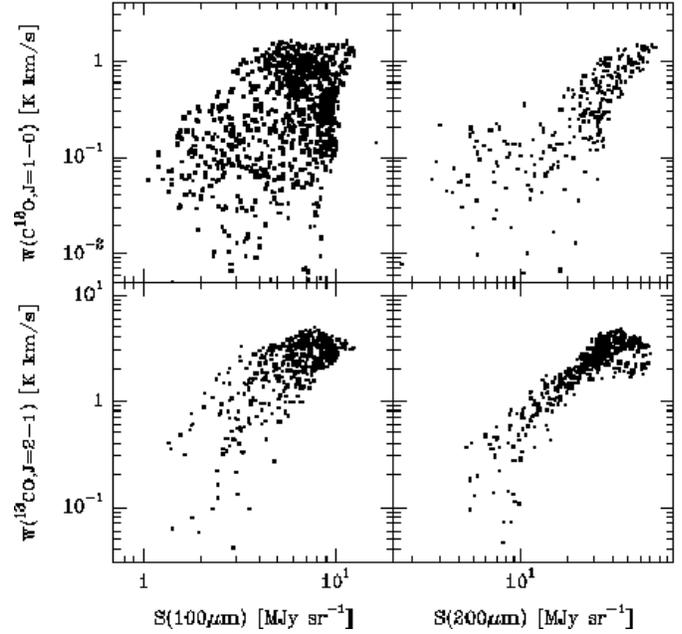}}
\caption[]{
C$^{18}$O(1--0) and $^{13}$CO(2--1) line areas as the function of 100$\mu$m
and 200$\mu$m surface brightness. In plots containing the 200\,$\mu$m surface
brightness values the original data are smoothed to resolution of
100$\arcsec$. In the case of 100\,$\mu$m the corresponding resolution is
70$\arcsec$ }
\label{fig:iso_w}
\end{figure}

In Fig.~\ref{fig:iso_w} C$^{18}$O(1--0) and $^{13}$CO(2--1) line areas are
compared with the 100\,$\mu$m and 200\,$\mu$m surface brightness values. Of
the two far-infrared bands 200\,$\mu$m is clearly better correlated with
molecular material. While on a logarithmic scale the relation between
C$^{18}$O(1--0) and 200\,$\mu$m emission is approximately linear, the
$^{13}$CO(2--1) intensity becomes saturated beyond
$S$(200\,$\mu$m)=50\,MJy\,sr$^{-1}$.
Based on IRAS observations Laureijs et al \cite{laureijs95} calculated the
difference $\Delta I_{100} = I(100\,\mu{\rm m}) - I(60\,\mu{\rm m})/0.21$. The
60\,$\mu$m emission traces the diffuse outer parts of the cloud and the
correlation with the $^{13}$CO core was found to be better for $\Delta I_{100}$
than for the original 100\,$\mu$m. The 100\,$\mu$m observations are affected
by this warmer emission and this causes the poor correlation. The colour
temperatures south of the L183 core could also be biased due to diffuse
material the presence of which is evident from the $^{12}$CO maps
(Fig.~\ref{fig:map_all}).

In Fig.~\ref{fig:colden_c18o_tau200}, the C$^{18}$O(1--0) column density (see
Fig.~\ref{fig:ph_colden} and Sect.~\ref{sect:model}) is compared with the dust
optical depth at 200\,$\mu$m. The figure confirms the good general correlation
between the two. The C$^{18}$O column density distribution may, however, be
slightly more extended in the SE. This was already seen in the C$^{18}$O(1--0)
distribution. On the other hand, 200\,$\mu$m optical depth is relatively larger
immediately south of the (0,0) position.

Nevertheless, the linear relation does hold only at lower column densities.
A linear least squares fit was made taking into account the uncertainty
in both variables. For points with $\tau(200\mu$m)$<$5$\cdot$10$^{-3}$ we get
\begin{equation}
\frac{N({\rm C}^{18}{\rm O})}{10^{13}\,{\rm cm}^{-2}}
= (4.62\pm0.16)10^{4} \, \tau(200\mu{\rm m}) - (1.5\pm1.0).
\label{eq:tau200_nc18o}
\end{equation}
Points at $\tau$(200$\mu$m)$<$5$\cdot$10$^{-5}$ were also discarded for this
fit since the C$^{18}$O column density estimates corresponding to these points
are no longer reliable. When $\tau(200\mu$m) exceeds $\sim$5$\cdot$10$^{-3}$
the C$^{18}$O column density is significantly below the prediction based on
lower optical depths. The C$^{18}$O emission could be reduced by depletion of
gas phase molecules, a large drop in the excitation temperature towards the
cold core or by the effect of optical depth.

\section{CO depletion in the cloud core}  \label{sect:depletion}

\begin{figure}
\resizebox{\hsize}{!}{\includegraphics{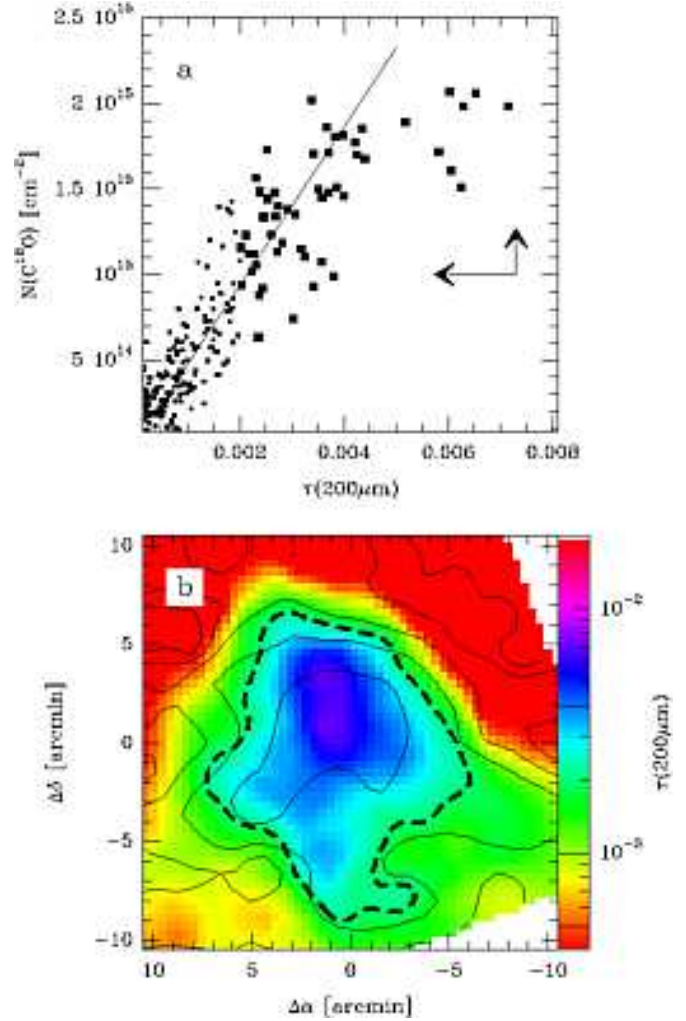}}
\caption[]{
{\bf a)} Correlation between the dust optical depth at 200\,$\mu$m and the C$^{18}$O
column density derived from C$^{18}$O(1--0) observations, assuming a constant
excitation temperature of 8.4\,K. The arrows indicate the shift if $N$(C$^{18}$O)
were calculated with a $\sim$2K lower excitation temperature, or if the
optical depth were calculated using a 0.5\,K higher dust temperature. The latter
applies to the highest $\tau(200\mu$m) values corresponding to $T_{\rm
dust}\sim$12\,K.
{\bf b)} The distribution of the 200\,$\mu$m optical depth over
the area of highest optical depth. The dotted contour indicates the level
$\tau(200\mu$m)=2$\cdot$10$^{-3}$. In (a) the points inside this contour
are plotted with larger symbols. The solid contours show the distribution of
the C$^{18}$O column density. The highest contour is at
1.6$\cdot$10$^{15}$\,cm$^{-3}$ and the others are each a factor of two below
the previous one. }
\label{fig:colden_c18o_tau200}
\end{figure}

The LTE analysis indicated C$^{18}$O(1--0) optical depths below one. In the
Monte Carlo model, the optical depth towards the cloud centre averaged over
the telescope beam was 0.56. These results were based on the $^{13}$CO(1--0)
and C$^{18}$O(1--0) observations. Since the emission of the observed lines may
come from different parts of the cloud, the determination of the optical depth
remains uncertain to some extent. In the positions close to the cloud centre
where C$^{18}$O(2--1) was observed LTE analysis based on the two C$^{18}$O
lines gave somewhat higher estimates for the optical depth, $\tau\la$1.0. 
The optical depth is insufficient to cause any significant saturation of the
C$^{18}$O line and will not affect the accuracy of the column density
estimates in Fig.~\ref{fig:colden_c18o_tau200}.

Based on the LTE analysis, the excitation temperature was rather uniform over
the cloud.
Compared with other parts of the cloud the excitation temperature towards
the cloud centre was lower by no more than 1\,K.
The column density estimates were based on an average value of 8.4\,K. In
Fig.~\ref{fig:colden_c18o_tau200}, the vertical arrow indicates the change in
the estimated column density if excitation temperature is reduced from a value
8.8\,K to a value of 7.0\,K. It is clear that even a drop of $\sim$2\,K is
insufficient to explain the relatively low C$^{18}$O column density estimates
at the position of high values of $\tau$(200$\mu$m).

The horizontal arrow in Fig.~\ref{fig:colden_c18o_tau200} indicates the shift
in the value of $\tau(200\mu$m) when a value of optical depth is calculated
with $T_{\rm dust}$=12.5\,K instead of 12.0\,K. Incorrect dust temperatures
are, however, not likely to explain the relative drop of the C$^{18}$O
intensity. Firstly, the error should be $\sim$1\,K or more at 12\,K to bring
the deviating points onto the fitted line even if the temperature were not
changed outside the core. Errors in temperatures could be caused either by
calibration or the background subtraction. If all 100\,$\mu$m surface
brightness values are multiplied by a constant, all $\tau(200\mu$m) estimates
would be scaled with a factor that has very weak temperature dependence. An
incorrect background subtraction, on the other hand, mostly affects the high
temperatures which correspond to low surface brightness values. As an example,
let us consider two dust temperatures, 12\,K and 15\,K. After the background
subtraction, in L183 these typically correspond to 100\,$\mu$m
surface brightness values of 7\,MJy\,sr$^{-1}$ and 3\,MJy\,sr$^{-1}$. If we
subtract 0.5\,MJy\,sr$^{-1}$ from the 100\,$\mu$m measurements the colour
temperatures would drop to 11.85\,K and 14.45\,K and the corresponding
200\,$\mu$m optical depth estimates would increase by 8\% and 20\%,
respectively. The net change is only some 12\% i.e. much less than the effect seen
in Fig.~\ref{fig:colden_c18o_tau200}.

Therefore, it seems that C$^{18}$O depletion remains the most probable cause
for the decreasing $N({\rm C^{18}O})/\tau(200\mu$m) ratio. 
In cloud cores the CO molecules freeze out onto the surfaces of the cold dust
grains. Direct evidence of the process is provided by the observations of the
molecular ice features (e.g. Tielens et al. \cite{tielens91}). As the result
the properties of the dust grains are altered and this may explain some
of the colour temperature variations observed in L183.
In the centre of L183 the visual extinction exceeds 10$^{\rm m}$ and CO
depletion has been detected in other clouds at similar extinction values.
Kramer et al. (\cite{kramer99}) and Bergin et al. (\cite{bergin01}) have
detected CO depletion in the cloud IC5146. In that cloud the relative
C$^{18}$O intensity decreases already below $A_{\rm V}$=10$^{\rm m}$.
Similarly in the cloud L977 a drop in C$^{18}$O intensity attributed to
depletion takes place at visual extinction $A_{\rm V}\sim10^{\rm m}$ (Alves et
al. \cite{alves99}). In L183 our models predicted central density close to
$10^5$\,cm$^{-3}$ for the spherical model with a radial density profile $\sim
r^{-1.5}$. This density is probably exceeded in the cold core and e.g. in
L1544 CO depletion was associated with densities above $10^5$\,cm$^{-3}$
(Caselli et al. \cite{caselli99}).

Based on Fig.~\ref{fig:colden_c18o_tau200}, the depletion factor is $\sim$1.5 in
the centre of L183 where the dust temperature is close to 12\,K. This is in
perfect agreement with the findings of Kramer et al. (\cite{kramer99}) who
have derived the depletion factor for IC5146 as a function of dust
temperature.

\section{DCO+ in the cold cloud core} \label{sect:dco}

H$^{13}$CO+ and DCO+ are concentrated close to the 200$\mu$m maximum. The
peaks of the intensity distributions of both species are very close to the
position of the continuum source, within $\sim 2\arcmin$. While the 200$\mu$m
emission continues to be strong towards the 100$\mu$m peak both H$^{13}$CO+
and DCO+ are restricted to a narrow ridge running in the north-south
direction. The distribution is therefore clearly different from either
C$^{18}$O or 200$\mu$m, although close to position (0,0) the 200$\mu$m
emission is also elongated in the north-south direction. More importantly, no
H$^{13}$CO+ or DCO+ emission is seen close to the 100$\mu$m maximum, while the
area of strong C$^{18}$O emission covers also that position.
% ##
At the position (4.5$\arcmin$, -2.5$\arcmin$) we get 2$\sigma$ upper limits of
0.10\,K for H$^{13}$CO+(1--0) and 0.20\,K for DCO+(2--1) main beam
temperature.

\begin{figure}
\resizebox{\hsize}{!}{\includegraphics{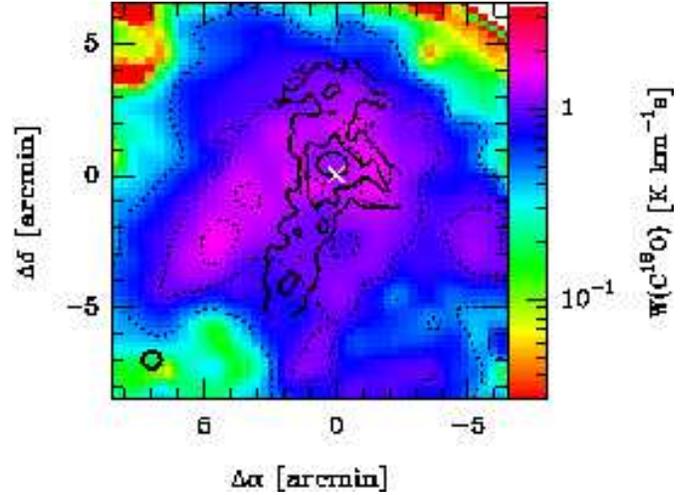}}
\caption[]{Contours of DCO+(2--1) integrated line area on the map
of integrated C$^{18}$O(1--0) line area. The DCO+ contours are at 0.2,
0.4, and 0.6\,K\,km\,s$^{-1}$. The dotted contours indicate C$^{18}$O(1--0)
intensity levels from 0.5 to 2.0\,K\,km\,s$^{-1}$ at intervals of
0.5\,K\,km\,s$^{-1}$.
The maps are presented at their original resolution. The beam sizes
(46$\arcsec$ for C$^{18}$O and 35$\arcsec$ for DCO+) are indicated with
circles at the lower left corner.} 
\label{fig:dco_c18o}
\end{figure}

Fig.~\ref{fig:dco_c18o} shows the DCO+ distribution in relation to C$^{18}$O.
Both peak in the same region around position (0,0) and the small extension
from this position towards the west can be interpreted to correspond to a similar
feature seen in C$^{18}$O. However, at the exact position of the DCO+ maximum,
the C$^{18}$O map shows a very slight depression, and in the south the two
species are clearly anticorrelated, with stronger C$^{18}$O emission seen on
each side of the DCO+ ridge.

The 100\,$\mu$m emission peaks some 5$\arcmin$ east of the DCO+ maximum and
even in details the distribution is completely unrelated to the DCO+ emission.
The correlation between DCO+ and 200\,$\mu$m emission is not very good either.
Both peak close to position (0,0) but compared with the FIR emission the DCO+
maximum is shifted one arcminute to the west.
Apart from a similar shift the correlation is good north of the centre
position. In the south the distributions of DCO+ and 200\,$\mu$m are
different. As already mentioned, no DCO+ emission was seen close to the
eastern FIR peak. The DCO+ map does not quite extend to the southern
200\,$\mu$m peak at position (1,-6) but
between the northern and southern 200\,$\mu$m peaks the DCO+ ridge goes over a
region with relatively low FIR emission.

\begin{figure}
\resizebox{\hsize}{!}{\includegraphics{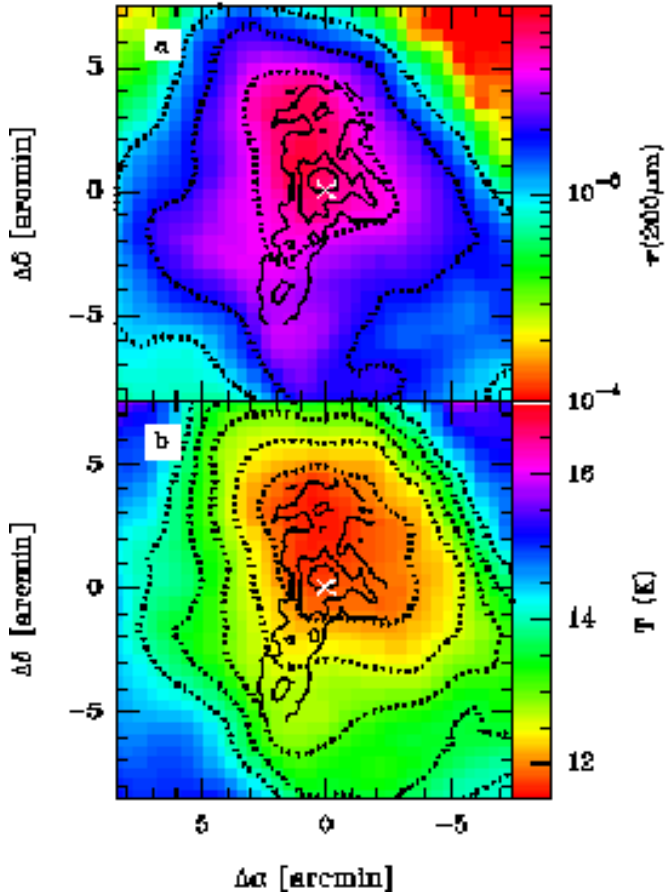}}
\caption[]{
{\bf a)} DCO+(2--1) contours on the map of 200\,$\mu$m optical depth. The
solid contours show the DCO+ distribution and are drawn at 0.2, 0.4, and
0.6\,K\,km\,s$^{-1}$. The dotted contours indicate values 1$\cdot 10^{-3}$,
2$\cdot 10^{-3}$ and 4$\cdot 10^{-3}$ of 200\,$\mu$m optical depth.
{\bf b)} DCO+(2--1) contours on the map of dust colour temperature. The solid DCO+
contours are drawn at 0.3, 0.5, 0.7 and 0.9\,K\,km$^{-1}$. The dotted contours
correspond to dust temperatures from 12 to 14\,K at steps of 0.5\,K.
The maps are presented at their original resolution.
}
\label{fig:dco_tau200_tdust}
\end{figure}

In Fig.~\ref{fig:dco_tau200_tdust}a we compare the DCO+ distribution with
the 200\,$\mu$m optical depth. Due to the dust temperature gradient,
$\tau(200\,\mu$m) is shifted relative to the 200\,$\mu$m emission and
coincides with the DCO+ distribution. Around position (0,0) and in the north,
the correlation between $\tau(200\,\mu$m) and DCO+ is very good. In the south,
the DCO+ follows roughly the dust optical depth distribution but the
correlation is not perfect. The $\tau(200\,\mu$m) distribution is more
extended SE of the centre position and the southern DCO+ peak at (1.5,\,~-4)
lies just north of the third $\tau(200\,\mu$m) peak, i.e. it does not follow
dust distribution. The southern part of the DCO+ ridge is outside the coldest
dust core (Fig.~\ref{fig:dco_tau200_tdust}b).

\begin{figure}
\resizebox{\hsize}{!}{\includegraphics{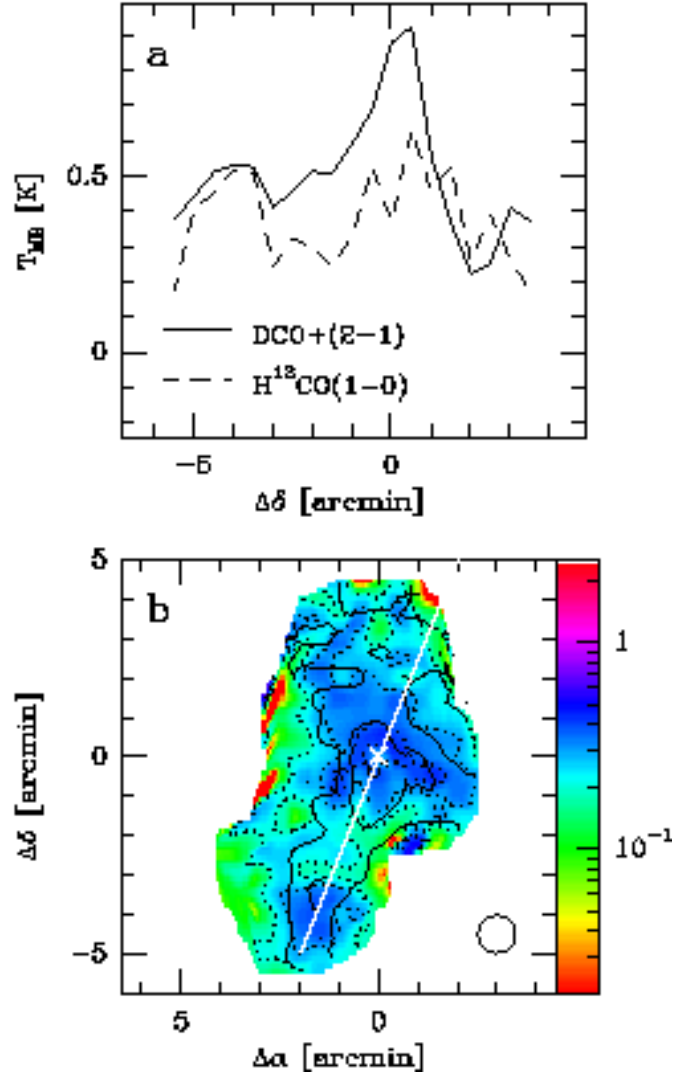}}
\caption[]{
{\bf a)} The DCO+(2--1) (solid line) and H$^{13}$CO+(1--0) (dashed line)
intensities along the line going through the two cores detected in these
lines. {\bf b)} DCO+(2--1) contours (solid lines) over-plotted on the
H$^{13}$CO+(1--0) map. The DCO+ contours levels are the same as in the
previous figures. The dotted lines indicate H$^{13}$CO+(1--0) values 0.2, 0.3
and 0.4\,K\,km\,s$^{-1}$. The DCO+ data are convolved to the resolution
of H$^{13}$CO+ observations. }
\label{fig:dco_h13co}
\end{figure}

As the distributions of the DCO+ and H$^{13}$CO+ are very similar, the same
conclusions apply to the comparison of the H$^{13}$CO+ distribution with CO
species and FIR emission. At the position (0,0), the observed main beam
temperatures are 1.5\,K and 0.79\,K for DCO+(2--1) and H$^{13}$CO+(1--0)
respectively. DCO+ seems to be more concentrated in this core, but the
difference is not significant when the difference in the beam size is taken
into account. In the upper panel of Fig.~\ref{fig:dco_h13co} the DCO+ and
H$^{13}$CO+ intensities are compared along the line going through the two
cores. The DCO+ values were convolved to the resolution of the H$^{13}$CO+
observations. In the main core, the H$^{13}$CO+ is displaced to the north with
respect to the DCO+. In the southern core, the relative intensity of
H$^{13}$CO+ is higher.

According to chemical models (Millar et al.\cite{millar89}; Roberts \& Millar
\cite{roberts00}) the ratio between
DCO+ and H$^{13}$CO+ can be used as a tracer of gas temperature. 
The reaction
\begin{equation}
H_3^{+} + HD \leftrightarrow H_2D^{+} + H_2 + \Delta E
\end{equation}
is strongly temperature dependent. At low temperatures the production of
$H_2D^{+}$ is enhanced and it reacts with CO to produce more DCO+. The
abundance ratio also depends, however, on other factors such as cloud age,
electron density and the total abundance of neutrals (Anderson et al.
\cite{anderson99}; Roberts \& Millar \cite{roberts00}). This prevents the use
of the ratio [DCO+]/[H$^{13}$CO+] for direct temperature determination.

The observed change in the intensity of the DCO+ and H$^{13}$CO+ lines
indicates a higher gas temperature in the southern core. This is consistent with
the difference in dust temperatures. Using the main beam antenna temperatures,
the ratio of DCO+(2--1) and H$^{13}$CO+(1--0) line areas increases from
$\sim$1.0 in the south to $\sim$1.7 in the main core. The excitation states of
DCO+ and H$^{13}$CO+ should be very similar. The isothermal Monte Carlo model
used to model the $^{13}$CO and C$^{18}$O lines was used to predict DCO+ lines
assuming an abundance of 5$\cdot$10$^{-11}$. Compared with the model in
Sect.~\ref{sect:model}, the density needed to be increased almost a factor of two to
produce lines $T_{\rm A}(J=1-0)\sim$\,0.8\,K, in which case the line ratio was
$T(1-0)/T(2-1)\sim$\,2. 
Assuming column density is directly proportional to observed intensity
and using a value of 66 for the ratio $^{12}$C/$^{13}$C (Langer \& Penzias
\cite{langer93}), we obtain values of 0.051 and 0.030 for the DCO+ to HCO+
abundance ratio in the two cores.
The ratios indicate gas temperatures $\sim$10\,K or less,
depending on the set of reaction constants used (Millar et al.\cite{millar89};
Roberts \& Millar \cite{roberts00}; see also Loren et al. \cite{loren90}). The
difference between the northern and southern cores corresponds to a
temperature change of at least 2\,K.

\section{Comparison of optical extinction with FIR dust and CO column densities}

\begin{figure}
\resizebox{8.0cm}{!}{\includegraphics{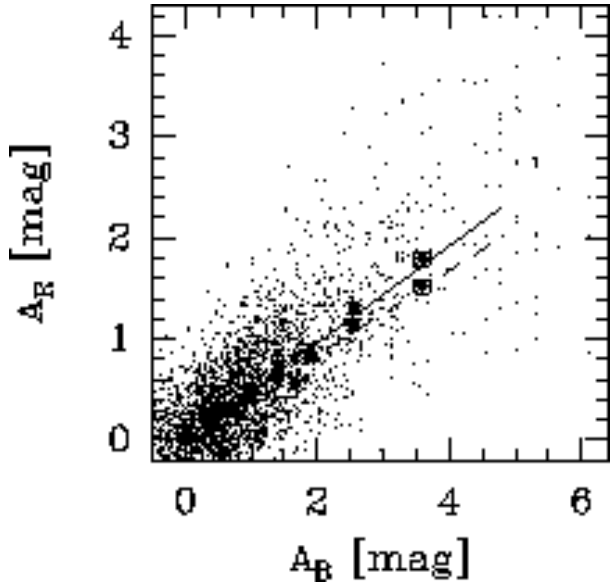}}
\caption[]{
The relation between $B$-band and $R$-band extinction as derived from the
starcounts. The lower points with errorbars are obtained by first averaging
the $R$-band counts in areas with given number of $B$-band stars. The dashed
line is the least squares line fitted to these. The upper points denote
averages of $A_{\rm R}$ for given $A_{\rm B}$ and the dots correspond to star
count averaged over circle with radius 1.7$\arcmin$. The solid line is the
least squares fit to these points and has a slope of 0.48}
\label{fig:starcount_starcount}
\end{figure}

In Fig.~\ref{fig:starcount_starcount} we plot $A_{\rm R}$ as a function
of $A_{\rm B}$, as derived from the starcounts (see Sect.~\ref{sect:counts}).
The dots denote individual extinction values obtained by averaging the counts
in each grid point over a circle with radius 1.7$\arcmin$. In the averaging
the weights were proportional to the area inside the circle which was,
however, only slightly larger than original 2.8$\arcmin\times$2.8$\arcmin$ square
used in the star counts.

We first calculated the average number of $R$-band stars in those
2.8$\arcmin\times$2.8$\arcmin$ squares with a given number of $B$-band stars. The
average counts in these areas were transformed into extinction values. The relation
with statistical errors is shown in Fig.~\ref{fig:starcount_starcount} (dashed
line). Linear least squares fit gives a relation $A_{\rm R}$ = (0.42 $\pm$ 0.02)
$A_{\rm B}$.
The fit takes into account uncertainty in both $A_{\rm B}$ and $A_{\rm R}$.
If the intercept were included in the fit its value would be only
$\sim0.03$.
If we calculate the average of the $A_{\rm R}$ values at all grid points with
a given $A_{\rm B}$ (instead of first calculating the average number of stars)
the derived relation becomes steeper, with a slope of 0.48$\pm$0.02. The value
is still somewhat lower than the value $\sim$0.56 for the extinction law with
$R_{\rm V}=A_{\rm B}/E_{\rm B-V}$=3.1. The fit is shown as a solid line in the
figure. The least squares line fitted directly to ($A_{\rm B}$, $A_{\rm R}$)
points gives also a slope of 0.48.
Considering the possible systematic errors of $\sim$10\% of our extinction
values, due to uncertainties in the calibration of $log N(m)$ vs. $m$ slopes,
our $A_{\rm R}/A_{\rm B}$ ratio is compatible with the standard reddening law
($R_{\rm V}=3.1$). On the other hand, it differs significantly from the value
of $A_{\rm R}/A_{\rm B}$=0.66 valid for the case of $R_{\rm V}=5.0$ as derived
for the ``outer-cloud dust'' (Mathis et al. \cite{mathis90}).

In the centre of L183 there were a few starcount squares where no stars were
seen and for Fig.~\ref{fig:starcount_starcount} the extinction was calculated
using a value of 0.5 stars per square. The values, $A_{\rm R}\approx$5$^{\rm
m}$, still only represent a lower limit of the true extinction. In the most
opaque area of 95 square arcminutes, four stars were detected in the $R$-band
and this translates into an extinction of 5.3$^{\rm m}$. In the $B$-band, no
stars were seen in this area while a single star would correspond to an
extinction of 6.8$^{\rm m}$. Therefore, it is likely that at the very centre
the extinction is at least $A_{\rm V}\sim$10$^{\rm m}$.

\begin{figure}
\resizebox{\hsize}{!}{\includegraphics{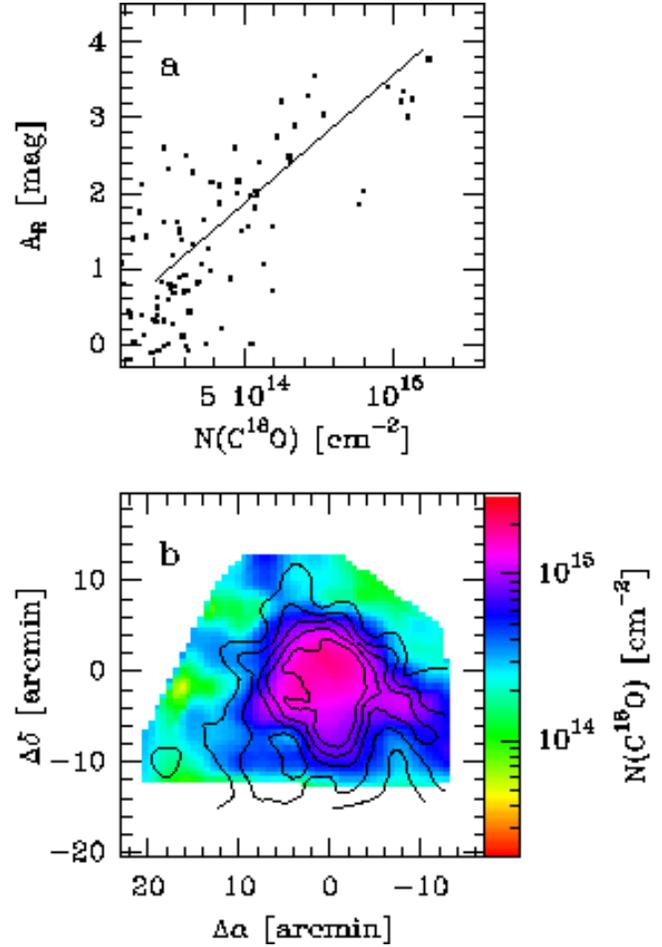}}
\caption[]{
{\bf a)} $R$-band extinction vs. C$^{18}$O column density.
{\bf b)} $R$-band extinction
contours drawn on the C$^{18}$O column density map. The contours of extinction
are drawn between 1$^{\rm m}$ and 5$^{\rm m}$ at intervals of 1$^{\rm m}$.
The C$^{18}$O column densities have been convolved to the resolution
of the extinction map, $\sim$2.8$\arcmin$.
}
\label{fig:extr_colden}
\end{figure}

\begin{figure}
\resizebox{\hsize}{!}{\includegraphics{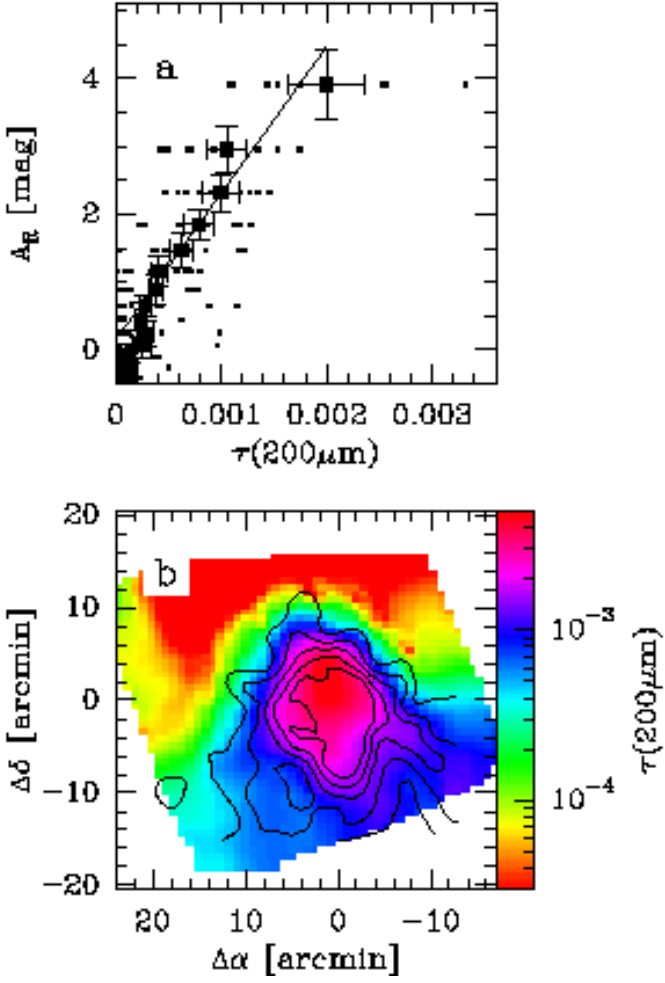}}
\caption[]{
{\bf a)} $R$-band extinction vs. 200\,$\mu$m optical depth. {\bf b)} $R$-band
extinction contours on the map of $\tau$(200$\mu$m) which has been
convolved to a resolution of 2.8$\arcmin$. The extinction contours are drawn
between 1$^{\rm m}$ and 5$^{\rm m}$ at intervals of 1$^{\rm m}$. }
\label{fig:extr_tau200}
\end{figure}

Next $A_{\rm R}$ is compared with the hydrogen column density estimate derived
from the C$^{18}$O(1--0) observations (Fig.~\ref{fig:ph_colden}) and the
200\,$\mu$m optical depth (Fig.~\ref{fig:c200_tau200}). 
The data are convolved to the resolution $2.8\arcmin$ corresponding to the
grid size used in the starcounts. The results are shown in
Figs.~\ref{fig:extr_colden} and \ref{fig:extr_tau200}.

The C$^{18}$O column density and the extinction are clearly correlated. The
C$^{18}$O depletion seen in Fig.~\ref{fig:colden_c18o_tau200} is not visible,
however, as the centre region where no $R$-band stars were detected is not
included in the figure. This limits the dynamical range available.
Furthermore, at low column densities the noise increases rapidly due to the
decreasing signal-to-noise ratio of the C$^{18}$O spectra. The maximum
extinction corresponding to one star per grid pixel area was $\sim$4$^{\rm
m}$. For the plot the star counts were again averaged over a circle with radius
1.7$\arcmin$ but points with $A_{\rm R}>$4$^{\rm m}$ are omitted in the
figure. The least squares line is
\begin{equation}
A_{\rm R}/{\rm mag}=(3.4\pm0.2) 10^{-15} N({\rm C}^{18}\rm{O})+(0.1\pm0.1).
\label{eq:nc18o_ar}
\end{equation}
With the earlier relation between $A_{\rm R}$ and $A_{\rm B}$ and the assumed
value of $R$=3.1 this translates to
$N$(C$^{18}$O)/cm$^{-2}$=1.9$\cdot$10$^{14}$A$_{\rm V}$-4$\cdot$10$^{13}$. The
slope is similar to the value found in regions of high extinction in Taurus
and Ophiuchus by Frerking et al. (\cite{frerking82}) but steeper than the
corresponding value for the lower range $2^{\rm m}<A_{\rm V}<4^{\rm m}$ where
the value in Taurus was 0.7$\cdot$10$^{14}$\,cm$^{-2}$\,mag$^{-1}$. The slopes
determined in L977 and IC5146 for $A_{\rm V}\la$10$^{\rm m}$ (Alves et al.
\cite{alves99}) are rather similar, $\sim$2$\cdot$10$^{14}$\,cm$^{-2}$\,mag$^{-1}$.
Also Harjunp\"a\"a et al. (\cite{harjunpaa96}) found a similar slope in Corona
Australis while in the Coalsack the value was only
$\sim$1.2$\cdot$10$^{14}$\,cm$^{-2}$\,mag$^{-1}$. Note that if we assume
standard extinction (Rieke \& Lebofsky \cite{rieke85}) to transform $A_{\rm
R}$ directly to $A_{\rm V}$ our slope increases to
2.2$\cdot$10$^{14}$\,cm$^{-2}$\,mag$^{-1}$.

Within the statistical uncertainty of the starcounts, the extinction map
agrees well with $\tau$(200$\mu$m). The dots in Fig.~\ref{fig:extr_tau200}
correspond to the star counts on the original grid. The filled squares with
error bars are averages for each discrete value of $A_{\rm R}$. The least
squares line based on the interval $1.0^{\rm m}<A_{\rm R}<4.0^{\rm m}$ is
\begin{equation}
A_{\rm R}/{\rm mag} = (2.2\pm0.3)10^3 \tau(200\mu{\rm m}) + (0.2\pm0.2).
\label{eq:tau200_ar}
\end{equation}
The relation can be used to derive an extrapolated extinction value for the
centre region where no stars were detected. In the centre there are several
200\,$\mu$m pixels with $\tau(200\mu{\rm m})\sim6 \cdot 10^{-3}$. According to
the previous relation, this corresponds to extinction $A_{\rm R}\approx 13^{\rm m}$
and visual extinction of the order of 17$^{\rm m}$. This clearly exceeds the
previous estimates, and is based on the sharp increase of the 200\,$\mu$m
optical depth in the cold cloud core.

Assuming a standard extinction law ($R_{\rm V}$=3.1) between $V$- and $R$-bands,
from Eq.~\ref{eq:tau200_ar} we obtain 
$\tau(200\mu{\rm m})/A_{\rm V}=3.4\cdot10^{-4}$\,mag$^{-1}$. 
For the cloud centre, the ratio increases to
3.8$\cdot10^{-4}$\,mag$^{-1}$ but is still below the value of
5.3$\cdot 10^{-4}$mag$^{-1}$ found in the Thumbprint Nebula (Lehtinen et
al.~\cite{lehtinen98}). In L183, the comparison between 100\,$\mu$m optical
depth and visual extinction gives a ratio 2.1$\cdot 10^{-3}$mag$^{-1}$ which
is similar to the value in the Thumbprint Nebula.

For a value of $\tau(200\mu{\rm m})=1.0\cdot 10^{-3}$, the previous relations
between $\tau(200\mu$m) and $A_{\rm R}$ and between $A_{\rm R}$ and C$^{18}$O
column density predict a column density of N(C$^{18}$O)=6.6$\cdot
10^{14}$\,cm$^{-2}$. Directly using the relation between $\tau(200\mu$m) and
N(C$^{18}$O) (Eq.~\ref{eq:tau200_nc18o}) we would obtain a lower value,
4.5$\cdot 10^{14}$\,cm$^{-2}$. One must remember, however, that
Eq.~\ref{eq:tau200_nc18o} was determined over a much larger range of optical
depths and consequently that relation may already be affected by the C$^{18}$O
depletion.

\section{Conclusions} \label{sect:conclusions}

The main conclusions drawn from the comparison of far-infrared data and
molecular line observations in L183 are:
\begin{itemize}
\item
The 100\,$\mu$m and 200\,$\mu$m emission clearly have different distributions.
\item
The 200\,$\mu$m maximum indicates the position of a core where either the dust
temperature becomes very low or the dust properties have changed. The colour
temperature determined from the 100\,$\mu$m and 200\,$\mu$m observations
with an emissivity law of $\nu^2$ is $\la$12\,K. The dust optical depth
peaks at the same position.
\item
Of the molecular lines, the optically thin C$^{18}$O best traces the dust
distribution. Although better correlated with the 200\,$\mu$m surface
brightness it shows strong emission at the locations of both the
100\,$\mu$m and 200\,$\mu$m maxima. The optically thick $^{12}$CO and
$^{13}$CO lines peak south of the dust cores. 
\item
C$^{18}$O is depleted in the cloud centre. The depletion factor is $\sim$1.5
in the core where dust temperature is close to 12\,K.
\item
Relative to each other, the DCO+ and H$^{13}$CO+ lines show very similar
distributions. Their distribution is completely different from any of the
observed CO lines but a good correlation does exist with the 200\,$\mu$m
optical depth.
\item
The estimated mass within 10$\arcmin$ of the (0,0) position was
25\,$M_{\sun}$, based on the 200\,$\mu$m optical depths and
$\sim$40\,$M_{\sun}$ based on the C$^{18}$O column
densities.
\item
The 100\,$\mu$m surface brightness is not well correlated with the extinction.
A linear relation does exist between $A_{\rm R}$ and dust optical depth at
200\,$\mu$m. Based on this correlation we derive for the cloud centre an
extinction $A_{\rm V}\sim 17^{\rm m}$.
\end{itemize}

\begin{acknowledgements}
Star counts were performed by J. Piironen (Helsinki University Observatory)
using observations made by O. Pizarro (ESO) on the request of G. Schnur
(Ruhr-Universit\"at Bochum) for this project. The ISOPHOT project was funded by
Deutsches Zentrum f\"ur Luft- und Raumfahrt (DLR), the
Max-Planck-Gesellschaft, the Danish, British and Spanish Space Agencies and
several European institutes. M.J., K.M., and K.L acknowledge the
support of the Academy of Finland Grant no. 1011055.
\end{acknowledgements}

\end{document}